\documentclass[pdflatex,sn-nature]{sn-jnl}

\usepackage{graphicx}
\usepackage{capt-of}
\usepackage{booktabs}
\usepackage{longtable}
\usepackage{array}
\usepackage{placeins}
\usepackage{flafter}
\usepackage{fancyhdr}
\usepackage{xcolor}
\usepackage{mdframed}
\usepackage{etoolbox}
\usepackage{titlesec}
\usepackage{amsmath}
\usepackage[cmintegrals,cmbraces]{newtxmath}
\PassOptionsToPackage{lining}{ebgaramond}
\usepackage{ebgaramond-maths}
\usepackage{microtype}
\usepackage{setspace}
\usepackage{xurl}
\usepackage{fvextra}
\usepackage{listings}
\usepackage[authorsoneline]{latexml}
\usepackage{fontspec}
\renewcommand{\arraystretch}{1.28}
\makeatletter
\renewcommand\normalsize{%
  \@setfontsize\normalsize{12bp}{18bp}%
  \abovedisplayskip 17pt plus 3pt minus 2pt%
  \abovedisplayshortskip 6pt plus 3pt%
  \belowdisplayshortskip 10pt plus 3pt minus 2pt%
  \belowdisplayskip 17pt plus 3pt minus 2pt%
  \let\@listi\@listI}
\renewcommand\small{%
  \@setfontsize\small{11bp}{15bp}%
  \abovedisplayskip 12pt plus 2pt minus 2pt%
  \abovedisplayshortskip 4pt plus 2pt%
  \belowdisplayshortskip 8pt plus 2pt minus 2pt%
  \belowdisplayskip 12pt plus 2pt minus 2pt%
  \let\@listi\@listI}
\renewcommand\footnotesize{\@setfontsize\footnotesize{10bp}{13bp}}
\renewcommand{\Titlefont}{\reset@font\bfseries\boldmath\fontsize{15bp}{20bp}\selectfont\centering}
\renewcommand{\Authorfont}{\reset@font\fontsize{11bp}{14bp}\selectfont\boldmath\centering}
\renewcommand{\addressfont}{\reset@font\fontsize{10.4bp}{12.8bp}\selectfont\centering\parskip=3pt}
\renewcommand{\abstractheadfont}{\reset@font\fontsize{14bp}{18bp}\bfseries\selectfont\raggedright}
\renewcommand{\abstractfont}{\reset@font\fontsize{12bp}{18bp}\selectfont\leftskip=0pt\rightskip=0pt\parfillskip=0pt plus 1fil}
\renewcommand{\figurecaptionfont}{\reset@font\fontfamily{\rmdefault}\fontsize{12bp}{18bp}\selectfont}
\renewcommand{\tablecaptionfont}{\reset@font\fontfamily{\rmdefault}\fontsize{12bp}{18bp}\selectfont}
\renewcommand{\tablebodyfont}{\reset@font\fontfamily{\rmdefault}\fontsize{11bp}{15bp}\selectfont}
\renewcommand{\tablecolheadfont}{\reset@font\fontfamily{\rmdefault}\fontsize{11bp}{15bp}\bfseries\boldmath\selectfont}
\renewcommand{\tablefootnotefont}{\reset@font\fontfamily{\rmdefault}\fontsize{10bp}{13bp}\selectfont}
\renewcommand{\sectionfont}{\reset@font\fontfamily{\rmdefault}\fontsize{17bp}{21bp}\bfseries\selectfont\boldmath\raggedright}
\renewcommand{\subsectionfont}{\reset@font\fontfamily{\rmdefault}\fontsize{14bp}{18bp}\bfseries\selectfont\boldmath\raggedright}
\renewcommand{\subsubsectionfont}{\reset@font\fontfamily{\rmdefault}\fontsize{13bp}{17bp}\bfseries\selectfont\boldmath\raggedright}
\renewcommand{\bmheadfont}{\reset@font\fontfamily{\rmdefault}\fontsize{13bp}{17bp}\bfseries\selectfont\boldmath\raggedright}
\long\def\abstract#1{\def\@abstract{%
  \begingroup
  \let\paragraph\subabstracthead
  \abstractfont
  \setlength{\parskip}{0pt}%
  {\abstractheadfont\abstractname\par}%
  \vspace{3pt}%
  #1\par
  \endgroup}}
\patchcmd{\@maketitle}{\newpage\null}{\newpage\vspace*{-46pt}}{}{}
\patchcmd{\@maketitle}{\else\vskip21pt\fi}{\else\vskip0pt\fi}{}{}
\patchcmd{\@maketitle}{\removelastskip\vskip20pt\nointerlineskip}{\removelastskip\vskip2pt\nointerlineskip}{}{}
\patchcmd{\@maketitle}{\vskip20pt}{\vskip8pt}{}{}
\patchcmd{\@maketitle}{\vskip7pt\addressfont}{\vskip5pt\addressfont}{}{}
\patchcmd{\@maketitle}{\removelastskip\vskip24pt}{\removelastskip\vskip0pt}{}{}
\patchcmd{\@maketitle}{\removelastskip\vskip24pt}{\removelastskip\vskip0pt}{}{}
\patchcmd{\@maketitle}{\removelastskip\vskip36pt}{\removelastskip\vskip0pt}{}{}
\makeatother
\titleformat{\section}[block]{\sectionfont}{}{0pt}{}
\titlespacing*{\section}{0pt}{23pt plus 4pt minus 3pt}{1pt}
\titleformat{\subsection}[block]{\subsectionfont}{}{0pt}{}
\titlespacing*{\subsection}{0pt}{18pt plus 3pt minus 2pt}{1pt}
\titleformat{\subsubsection}[block]{\subsubsectionfont}{}{0pt}{}
\titlespacing*{\subsubsection}{0pt}{17pt plus 3pt minus 2pt}{1pt}
\renewcommand{\bmhead}[1]{%
  \par\addvspace{16pt plus 3pt minus 2pt}%
  \noindent{\bmheadfont #1\par}\vspace{1pt}}
\normalsize
\bibpunct{}{}{,}{s}{}{,}

\fancypagestyle{manuscript}{%
  \fancyhf{}
  \fancyfoot[R]{\fontsize{11bp}{13bp}\selectfont\thepage}
  
  }
\fancypagestyle{firstpage}{%
  \fancyhf{}
  \fancyfoot[R]{\fontsize{11bp}{13bp}\selectfont\thepage}
  
  }
\definecolor{manuscriptlink}{HTML}{264A6B}
\definecolor{quotebg}{HTML}{F3F6F8}
\definecolor{quoterule}{HTML}{355B78}
\definecolor{sourceframe}{HTML}{B9C4CC}
\definecolor{sourcebg}{HTML}{F8FAFB}
\AtBeginDocument{\hypersetup{colorlinks=true,linkcolor=black,citecolor=black,urlcolor=manuscriptlink}}

\renewenvironment{quote}{%
  \begin{mdframed}[
    backgroundcolor=quotebg,
    linecolor=quoterule,
    linewidth=0pt,
    leftline=true,
    leftmargin=0pt,
    rightmargin=0pt,
    innerleftmargin=14pt,
    innerrightmargin=14pt,
    innertopmargin=11pt,
    innerbottommargin=11pt,
    skipabove=12pt,
    skipbelow=14pt,
    splittopskip=8pt,
    splitbottomskip=8pt]
  \normalfont\normalsize\setlength{\parindent}{0pt}\setlength{\parskip}{7pt}%
}{\end{mdframed}}
\newcommand{\agentquoteopen}{\iflatexml\else``\fi}
\newcommand{\agentquoteclose}{''}

\newcommand{\titlecorrespondence}{%
  \par\vspace{-4pt}%
  \begingroup\centering
  \fontsize{11bp}{14bp}\selectfont
  *Corresponding authors: \href{mailto:arao@broadinstitute.org}{arao@broadinstitute.org},
  \href{mailto:eric@broadinstitute.org}{eric@broadinstitute.org},
  \href{mailto:pardis@broadinstitute.org}{pardis@broadinstitute.org}
  \par\endgroup
  \vspace{8pt}}
\let\originalprintabstract\printabstract
\renewcommand{\printabstract}{%
  \titlecorrespondence
  \originalprintabstract
  }

\newcommand{\sourcefigure}[2][]{\includegraphics[pagebox=artbox,#1]{#2}}
\newcommand{\sourcepanel}[3][]{\includegraphics[pagebox=artbox,trim=#3,clip,#1]{#2}}

\newcolumntype{L}[1]{>{\raggedright\arraybackslash}p{#1}}
\newcommand{\sourcebaselinestretch}{1}
\newcommand{\supplementarysource}[4][\clearpage]{%
  #1
  \subsection{#2}
  \begingroup
  \small\color{manuscriptlink}
  \noindent\texttt{\detokenize{#3}}\par
  \endgroup
  \vspace{\parskip}
  \iflatexml
  \lstinputlisting[
    basicstyle=\ttfamily\footnotesize,
    columns=fullflexible,
    keepspaces=true,
    showstringspaces=false,
    breaklines=true,
    frame=single
  ]{#4}%
  \else
  \VerbatimInput[
    fontsize=\footnotesize,
    baselinestretch=\sourcebaselinestretch,
    breaklines=true,
    breakanywhere=true,
    breakanywheresymbolpre={},
    breakanywheresymbolpost={},
    breaksymbolleft={},
    breaksymbolright={},
    tabsize=4,
    frame=single,
    rulecolor=\color{sourceframe},
    fillcolor=\color{sourcebg},
    framesep=8pt
  ]{#4}%
  \fi}

\begin{document}

\title[Science sandboxes measure scientific capability]{Science sandboxes measure the scientific capability of AI agents}

\iflatexml
\author{\fnm{Arya S.} \sur{Rao}\textsuperscript{1,2,*}}
\author{\fnm{Rodrigo I.} \sur{Castro}\textsuperscript{3}}
\author{\fnm{Sager J.} \sur{Gosai}\textsuperscript{4}}
\author{\fnm{Kenneth B.} \sur{Hsu}\textsuperscript{1,10}}
\author{\fnm{Yasha} \sur{Ektefaie}\textsuperscript{1}}
\author{\fnm{Shantanu} \sur{Singh}\textsuperscript{1}}
\author{\fnm{Sangeeta N.} \sur{Bhatia}\textsuperscript{1,5,6,7,8}}
\author{\fnm{Steven K.} \sur{Reilly}\textsuperscript{9}}
\author{\fnm{Ryan} \sur{Tewhey}\textsuperscript{3}}
\author{\fnm{Eric S.} \sur{Lander}\textsuperscript{1,10,11,*}}
\author{\fnm{Pardis C.} \sur{Sabeti}\textsuperscript{1,6,12,13,*}}
\date{\footnotesize
\textsuperscript{1}The Broad Institute of MIT and Harvard; Cambridge, MA 02142, USA\\
\textsuperscript{2}Department of Biomedical Informatics, Harvard Medical School; Boston, MA 02115, USA\\
\textsuperscript{3}The Jackson Laboratory; Bar Harbor, ME 04609, USA\\
\textsuperscript{4}Sutter Hill Ventures; Palo Alto, CA 94304, USA\\
\textsuperscript{5}David H. Koch Institute for Integrative Cancer Research, Massachusetts Institute of Technology, Cambridge, MA 02139, USA\\
\textsuperscript{6}Howard Hughes Medical Institute, Chevy Chase, MD 20815, USA\\
\textsuperscript{7}The Wyss Institute for Biologically Inspired Engineering at Harvard University, Boston, MA 02115, USA\\
\textsuperscript{8}Harvard-MIT Program in Health Sciences and Technology, Institute for Medical Engineering and Science, Massachusetts Institute of Technology, Cambridge, MA 02139, USA\\
\textsuperscript{9}Department of Genetics, Yale School of Medicine; New Haven, CT, USA\\
\textsuperscript{10}Department of Systems Biology, Harvard Medical School, Boston, MA, USA\\
\textsuperscript{11}Department of Biology, Massachusetts Institute of Technology, Cambridge, MA, USA\\
\textsuperscript{12}Department of Immunology and Infectious Diseases, Harvard T.H. Chan School of Public Health, Boston, MA 02115, USA\\
\textsuperscript{13}Department of Organismic and Evolutionary Biology, Harvard University, Cambridge, MA 02138, USA\\[4pt]
*Corresponding authors: \href{mailto:arao@broadinstitute.org}{arao@broadinstitute.org},
\href{mailto:eric@broadinstitute.org}{eric@broadinstitute.org},
\href{mailto:pardis@broadinstitute.org}{pardis@broadinstitute.org}}
\else
\author*[1,2]{\fnm{Arya S.} \sur{Rao}}
\author[3]{\fnm{Rodrigo I.} \sur{Castro}}
\author[4]{\fnm{Sager J.} \sur{Gosai}}
\author[1,10]{\fnm{Kenneth B.} \sur{Hsu}}
\author[1]{\fnm{Yasha} \sur{Ektefaie}}
\author[1]{\fnm{Shantanu} \sur{Singh}}
\author[1,5,6,7,8]{\fnm{Sangeeta N.} \sur{Bhatia}}
\author[9]{\fnm{Steven K.} \sur{Reilly}}
\author[3]{\fnm{Ryan} \sur{Tewhey}}
\author*[1,10,11]{\fnm{Eric S.} \sur{Lander}}
\author*[1,6,12,13]{\fnm{Pardis C.} \sur{Sabeti}}

\affil*[1]{\orgname{The Broad Institute of MIT and Harvard; Cambridge, MA 02142, USA}}
\affil[2]{\orgname{Department of Biomedical Informatics, Harvard Medical School; Boston, MA 02115, USA}}
\affil[3]{\orgname{The Jackson Laboratory; Bar Harbor, ME 04609, USA}}
\affil[4]{\orgname{Sutter Hill Ventures; Palo Alto, CA 94304, USA}}
\affil[5]{\orgname{David H. Koch Institute for Integrative Cancer Research, Massachusetts Institute of Technology, Cambridge, MA 02139, USA}}
\affil[6]{\orgname{Howard Hughes Medical Institute, Chevy Chase, MD 20815, USA}}
\affil[7]{\orgname{The Wyss Institute for Biologically Inspired Engineering at Harvard University, Boston, MA 02115, USA}}
\affil[8]{\orgname{Harvard-MIT Program in Health Sciences and Technology, Institute for Medical Engineering and Science, Massachusetts Institute of Technology, Cambridge, MA 02139, USA}}
\affil[9]{\orgname{Department of Genetics, Yale School of Medicine; New Haven, CT, USA}}
\affil[10]{\orgname{Department of Systems Biology, Harvard Medical School, Boston, MA, USA}}
\affil[11]{\orgname{Department of Biology, Massachusetts Institute of Technology, Cambridge, MA, USA}}
\affil[12]{\orgname{Department of Immunology and Infectious Diseases, Harvard T.H. Chan School of Public Health, Boston, MA 02115, USA}}
\affil[13]{\orgname{Department of Organismic and Evolutionary Biology, Harvard University, Cambridge, MA 02138, USA}}
\fi

\abstract{Scientific progress depends not only on finding solutions, but on learning the rules that explain why they work and using that understanding to design better experiments. We introduce science sandboxes, a framework for studying this capability in AI agents through repeated cycles of experimentation, feedback, and hypothesis revision. Science sandboxes invite an agent to query the natural world in different ways, ranging from ``wet'' physical experiments, to ``damp'' predictive models trained on empirical data, to ``dry'' invented rules. By establishing a common experimental loop and a protocol for evaluating agents within it, science sandboxes allow assessment of both quantitative performance on specific metrics and qualitative scientific reasoning, across a spectrum of empirical verifiability. Here, we instantiate this framework in two biological settings, models of regulatory genomics and protein fitness prediction, and examine the capabilities of frontier agents. Across these settings, we could see when agents successfully optimized a quantitative metric without understanding the rules underlying the system. In particular, their scientific reasoning deteriorated when they encountered systems whose rules fell outside familiar biological priors. By highlighting such failure modes, science sandboxes make the frontier of scientific capability measurable and provide a controlled setting in which to study and ultimately expand it.}

\maketitle
\thispagestyle{firstpage}
\pagestyle{manuscript}

\clearpage
\section{Introduction}\label{introduction}

Experimental science advances by asking questions of nature to learn its rules. Consider the problem of learning how DNA sequences control gene expression. An investigator might select a set of sequences, measure their activity in the laboratory, infer which features of the sequences matter, and use that hypothesis to design a more informative set. In the course of this work, the aim is not merely to find sequences that drive gene expression, but to discover and understand regularities that explain why they work and guide the next experiment. A true measure of scientific capability should therefore assess whether experimental evidence leads to better hypotheses and more informative experiments, reflecting growing understanding rather than optimization alone. If artificial intelligence (AI) is to become a useful partner in scientific discovery, it must demonstrate this ability, motivating growing efforts to develop and evaluate increasingly capable scientific AI systems.\cite{Gao2024-ar,Gottweis2026-ts,Swanson2025-bw,Ghareeb2026-dh,Majumder2024-du,Laurent2024-xu,Starace2025-ah,Mitchener2025-yn}

We introduce the science sandbox: a controlled testbed for measuring scientific capability. A science sandbox recreates the essential structure of science by allowing an autonomous investigator to conduct successive experiments, interpret the results, and use what it learns to refine its hypotheses and choose the next experiments. Because the source of experimental feedback is hidden, the investigator must infer the rules of the system from the evidence alone. This source may be a physical experiment, a computational model of one, or an invented system whose rules are known only to the designers of the sandbox. By preserving the investigator's evolving hypotheses and experimental choices, a sandbox reveals whether improved performance simply reflects mathematical optimization or is based on a deeper understanding of the system. Traditional benchmarks cannot make this distinction: they reward any strategy that raises the score.

Here, we describe science sandboxes for two archetypal problems in biology, evaluating the scientific capabilities of frontier agentic AI systems. The first sandbox, MPRAbox, instantiates the sequence-selection problem posed above: each agent is tasked with selecting maximally informative libraries of regulatory DNA sequences for training downstream predictive models, a task of choosing a large experimental set from a combinatorially vast sequence space. The second sandbox, CodonBox, tests agents' ability to conduct \emph{de novo} rule discovery, asking, as early molecular geneticists once did, how an unfamiliar genetic system translates sequence into function.\cite{Nirenberg1961-qu,Crick1961-wg} Together, these sandboxes test AI as a scientific investigator, not merely as a solver of scientific tasks. In the process, they open a path towards understanding and scaling scientific capability itself.

\clearpage
\section{Results}\label{results}

\subsection{Science sandboxes measure how agents learn from experiments}

A science sandbox is a controlled environment in which AI agents can conduct experiments to learn the rules governing a phenomenon (Figure 1a). In each round, the agent chooses what to test based on its initial hypotheses, receives feedback from a sealed oracle, revises its hypotheses based on the feedback, and selects the next experiment. The aim is not only to determine whether the agent can improve a score, but whether it can behave like an experimental scientist: proposing informative tests, interpreting evidence, and converging toward coherent rules.

\begin{figure}[!htbp]
\centering
\iflatexml
\includegraphics[width=800pt]{figures/html-svg/figure1.svg}
\else
\sourcefigure[width=0.985\textwidth,height=0.69\textheight,keepaspectratio]{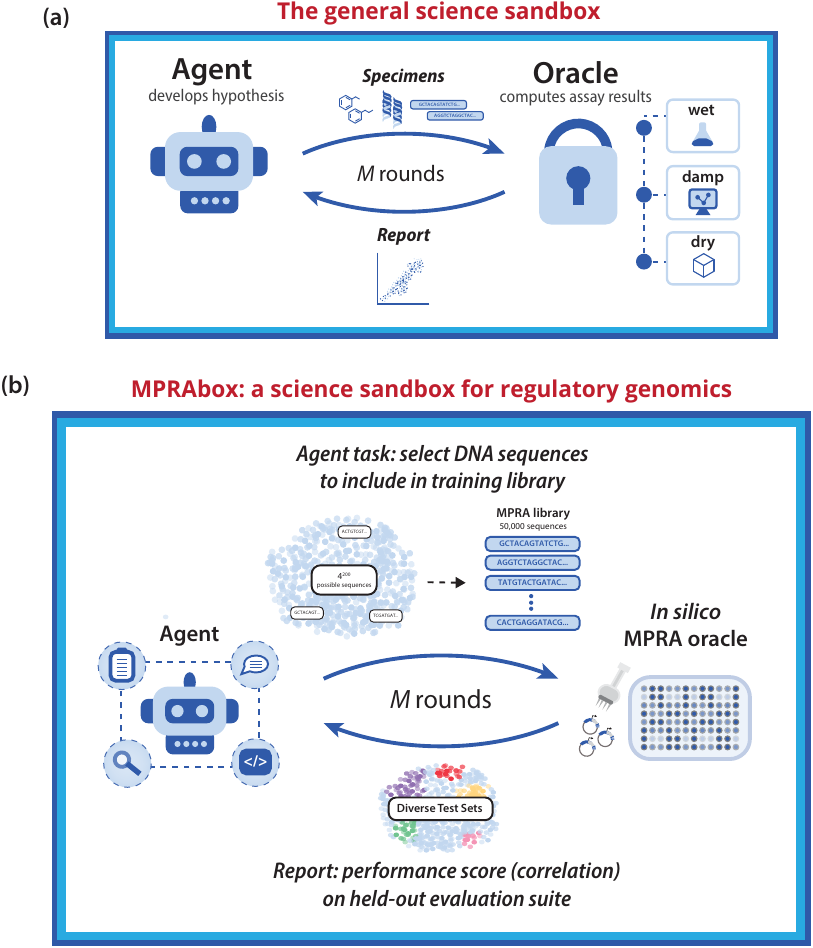}
\fi
\caption{\textbf{Science sandboxes for closed-loop evaluation of AI experimentalists.} (a) A science sandbox couples an autonomous agent to a sealed oracle through a closed loop of action, feedback, and hypothesis revision. At each design round, the agent proposes a batch of actions, receives feedback from the oracle, records its interpretation, and selects its next actions. The oracle can be wet (a physical experiment), damp (a model trained on experimental data), or dry (an invented rule created by the experimenter). (b) MPRAbox instantiates this framework in regulatory genomics. The agent selects a finite massively parallel reporter assay (MPRA) library from the effectively unbounded space of possible 200 bp regulatory DNA sequences. The library is evaluated by a sealed \emph{in silico} MPRA oracle that acts as a simulated wet lab, and aggregate performance metrics from diverse hidden evaluation sets are returned to guide subsequent design rounds.}
\label{fig:science-sandboxes}
\end{figure}

Every science sandbox has three core parts: \emph{specimens}, \emph{assays}, and an \emph{oracle}. By specimens, we broadly mean any entity that can be submitted for testing, such as a DNA sequence, protein, chemical, cell, designed construct, or abstract string. By assays, we mean methods for ascertaining specific properties of those specimens. By an oracle, we mean a hidden mechanism that applies the assays to each specimen, analyzes the results, and returns a report to the agent, which may contain the full results or just a limited summary. In general, sandboxes can contain multiple kinds of specimens and multiple assays, and the oracle's reports can range from a single scalar value to rich outputs such as images, time series, or molecular structures. For simplicity, we focus here on sandboxes with one type of specimen, one assay, and reports consisting of a handful of numbers.

Oracles can vary in how they apply assays to generate feedback on chosen specimens. \emph{Wet oracles} obtain results from actual physical experiments. \emph{Damp oracle}s use computational models trained on empirical data to approximate experimental results. \emph{Dry oracles} apply invented rules specified by the sandbox designer, which may have no relationship to the natural world. Each type serves a different purpose. Wet oracles provide realism, damp oracles make repeated experimentation scalable, and dry oracles can test how well agents can infer hidden rules in arbitrary settings.

A sandbox may have a known physical meaning, a concealed physical meaning, or no physical meaning at all. In the first case, the sandbox corresponds to real biological or chemical systems and the agents are informed of this fact, allowing them to draw on published scientific knowledge in deciding what to test. In the second case, the sandbox again corresponds to real biological or chemical systems, but the information is made abstract, for example, by replacing DNA sequences (strings of \(A,C,G,T\)) with arbitrary character strings (e.g., from \(\{0,1,2,3\}\) or \(\{\#,\$,\%,\&\}\)), with the goal of preventing agents from relying on familiar biological priors. This distinction is useful because it allows a sandbox to ask not only whether an agent can exploit known science, but whether it can infer unfamiliar rules from evidence alone.

The central object of evaluation is the agent's reasoning, not simply its final score. A science sandbox asks agents to conduct a series of experiments and to record their reasoning in a lab notebook. These notebooks allow humans, or independent AI judges, to evaluate whether the agent merely found higher-scoring specimens through local search or instead formed, tested, and revised hypotheses that explain the system. In this sense, a science sandbox is not just a benchmark of performance. It is a controlled setting for measuring scientific exploration itself.

This framework is intentionally broad. Some biological examples include:

(1) Cell Growth Assay, with (i) the specimens being a collection of chemicals and (ii) the assay being applying a specific concentration of a molecule to a specific number of cells and determining percentage change in the number of live cells present five days later. The report might consist of the percentage change observed for each chemical tested.

(2) Protein Structure, with (i) the specimens being amino acid sequences and (ii) the assay being a determination of the structure of the corresponding protein. The report might contain only the proportion of amino-acid residues in an alpha-helical conformation.

(3) Massively Parallel Reporter Assay (MPRA), with (i) the specimens being a DNA sequence of length 200 (``candidate enhancer'') and (ii) the assay being inserting a candidate enhancer upstream of the promoter of a particular gene in a cell type and measuring the proportional change in transcription level (amount of RNA produced). The report might consist of the change observed for each of the candidate enhancers submitted for testing or, alternatively, just the mean change across all these candidate enhancers.

We will begin by studying this last example.

\subsection{MPRAbox is a science sandbox for regulatory sequence design}\label{mprabox-is-a-science-sandbox-for-regulatory-sequence-design}

Regulatory genomics ---~specifically, the example of MPRA noted above ---~provides a natural setting for a science sandbox, because it combines (i) a combinatorially vast space of DNA sequences, providing ample specimens, with (ii) an assay that can be experimentally performed at large scale, effectively approximated by computational models at even greater scale, or scored by a rich array of arbitrary rules (Figure 1b). The fundamental goal of MPRA is to understand how regulatory DNA sequences drive gene expression.\cite{Tewhey2016-rx,Kircher2019-hl,Siraj2026-na}

In the simplest form of the MPRA sandbox, an oracle would receive from an agent a ``library'' of \(N\) candidate enhancers and return to the agent a report containing the results (proportional change in expression) for each of the \(N\) candidate enhancers. Those results could come directly from a laboratory experiment or, for faster iteration, from a computational model that approximates the assay. For example, one could use Malinois, a published model trained on more than 700,000 experimental MPRA measurements widely used across academia and industry to approximate wet-lab results.\cite{Gosai2024-lx,noauthor_undated-hp,noauthor_undated-iv,Butts2025-he,Ghari2025-ih,Ma2026-ar,Weykopf2026-or,Uehara2025-wj} Based on the report, the agent could select a new set of specimens that it believes will have higher scores.

A more interesting and useful challenge is to ask agents not simply to find high-scoring enhancers, but to design the most informative library of size \(N\) for training a predictive sequence-to-activity model. Specifically, researchers use MPRA data to train predictive sequence-to-activity models and, because the amount of data they can collect is limited, they want to use the \emph{most informative} set of candidate enhancers for training the model --- that is, the library of \(N\) specimens that will most improve training of a predictive sequence-to-activity model. We wanted to see how agents perform on this task, allowing them to submit distinct libraries of \(N\) candidate enhancers across one or more rounds (Figure 1b).

In this version of the MPRA sandbox, the oracle returns to the agent only a short report, consisting of a handful of scalar values that summarize the value of the library in training a model (Figure 1b). Specifically, we use a damp oracle that performs the following steps: (1) for each of the \(N\) candidate enhancers, calculate the results of the MPRA assay by using Malinois; (2) based on the (candidate enhancer, result) pairs from Malinois, train a new sequence-to-activity model from scratch; (3) evaluate the performance of the new model on 14 hidden test sets of candidate enhancers, consisting of 5 sets for which the activity scores are based on empirical laboratory measurements and 9 sets for which the activity scores are based on the Malinois model (Table 1); and (4) return to the agent a summary report consisting of only (i) the Pearson correlation between the new model's prediction and the ``ground truth'' for each of the 14 sets and (ii) an overall performance score, consisting of the mean of these 14 numbers. (The agents receive no other information, including about the nature or contents of the 14 sets.)

On each round, agents received an instructions.md file (Supplementary Information) directing them to design an MPRA library for training a generalizable sequence-to-activity model and to maintain an ongoing ``lab notebook'' documenting their reasoning. The task was not only to design a high-performing library, but also to develop a theory of what makes a library informative and use successive rounds to test and revise that theory.

We set \(N = 50{,}000\), because this value is typical for MPRA libraries, is large enough to train meaningful predictive models, and is small enough that the constraint on library size is a major determinant of performance (Supplementary Figure 1).

As noted above, our goal is not only to assess the agent's quantitative performance (correlation score), but also to probe the agent\textquotesingle s reasoning to discover rules that explain why some libraries are more informative than others.

\begin{table}[!htbp]
\caption{\textbf{Hidden test sets used in MPRAbox.} We evaluated models across nine sequence collections representing experimentally measured regulatory sequences, human genetic variants, annotated regulatory elements, genomic background, and synthetic DNA. For five collections, we evaluated predictions against both wet-lab MPRA measurements and Malinois-predicted activity, yielding 14 labeled evaluation sets in total. All genomic sequences came from chromosomes excluded from Malinois training. Full construction and filtering criteria are described in Methods.}
\label{tab:evaluation-sets}
\centering
\small
\begin{tabular}{@{}L{0.19\textwidth}L{0.39\textwidth}L{0.22\textwidth}r@{}}
\toprule
\textbf{Evaluation set} & \textbf{Sequence source} & \textbf{Labels} & \(n\) \\
\midrule
MPRA holdout, chr7/13 & Held-out sequences from the Gosai et al. episomal MPRA & Experimental + Malinois & 60,055 \\
MPRA holdout, chr19/21/X & Independent held-out sequences from the Gosai et al. episomal MPRA & Experimental + Malinois & 56,340 \\
UKBB/GTEx fine-mapped & Fine-mapped UK Biobank and GTEx cis-eQTL variants & Experimental + Malinois & 59,084 \\
UKBB/GTEx, both alleles & UK Biobank and GTEx variants with reference and alternate alleles retained & Experimental + Malinois & 62,966 \\
UKBB/GTEx, one allele & UK Biobank and GTEx variants represented by one allele per locus & Experimental + Malinois & 30,505 \\
Sei classes & Genomic regions spanning Sei sequence classes & Malinois & 20,000 \\
DHS index & DNase I hypersensitive sites & Malinois & 20,000 \\
Genomic windows & Random genomic windows & Malinois & 20,000 \\
Synthetic DNA & Uniformly random 200-bp sequences & Malinois & 20,000 \\
\bottomrule
\end{tabular}
\end{table}

\clearpage
\begingroup
\fontsize{10.5bp}{13.2bp}\selectfont
\renewcommand{\arraystretch}{1.05}
\begin{longtable}{@{}L{0.26\textwidth}L{0.69\textwidth}@{}}
\caption{\textbf{Human-selected baseline strategies.} Strategy names correspond directly to the labels in Fig. 2. DHS denotes a DNase I hypersensitive site, a genomic region of accessible chromatin. NMF denotes non-negative matrix factorization, which we used to summarize variation in DHS activity across cell types into 16 recurring regulatory patterns, referred to here as topics. Sei is a sequence-based regulatory model that assigns genomic sequences to regulatory classes. Oracle labels are activity values predicted by Malinois; real MPRA labels are experimentally measured activities. The mixed strategies combine sequences selected by these individual approaches in the proportions indicated.}\label{tab:baseline-strategies}\\
\toprule
\textbf{Strategy} & \textbf{Description} \\
\midrule
\endfirsthead
\multicolumn{2}{c}{\tablename\ \thetable\ (continued)}\\
\toprule
\textbf{Strategy} & \textbf{Description} \\
\midrule
\endhead
\midrule
\multicolumn{2}{r}{Continued on next page}\\
\endfoot
\bottomrule
\endlastfoot
DHS topic-weighted & Sequences were drawn from DNase I hypersensitive sites (DHSs), genomic regions associated with accessible chromatin. We used non-negative matrix factorization (NMF) to represent patterns of DHS activity across cell types as 16 regulatory ``topics.'' DHSs were sampled with probability proportional to their loading on these NMF topics. \\
DHS random & Sequences were drawn uniformly at random from the same collection of DHSs, without using the NMF topics to determine sampling frequency. \\
DHS component-stratified & DHSs were grouped according to the 16 NMF regulatory topics, and the library allocated an equal number of sequences to each topic rather than sampling according to their natural abundance. \\
SEI class-balanced & Sequences were drawn from genomic regions assigned to regulatory classes by Sei, a sequence-based model that classifies regulatory activity. Sampling was based on the Sei regulatory classes rather than directly on DHS activity. \\
SEI random & Sequences were drawn uniformly at random from the same Sei-annotated genomic regions, without balancing or weighting by regulatory class. \\
Random synthetic + oracle labels & Each 200-bp sequence was generated synthetically by choosing \(A\), \(C\), \(G\), or \(T\) independently with equal probability at every position. Activity labels were then assigned by the Malinois oracle rather than measured experimentally. \\
Malinois training sequences + oracle labels & Sequences were sampled uniformly from the original Malinois training set, and their activity labels were generated by the Malinois oracle. \\
Malinois training sequences + real MPRA labels & The same type of sequences were sampled from the original Malinois training set, but the labels were the experimentally measured MPRA activities used to train Malinois rather than Malinois predictions. \\
\textonehalf{} DHS + \textonehalf{} SEI & Half of the library was drawn using the DHS topic-weighted strategy and half using the Sei class-based strategy. \\
\textonehalf{} DHS + \textonehalf{} synth & Half of the library was drawn using the DHS topic-weighted strategy and half consisted of uniformly random synthetic DNA sequences. \\
DHS stratified + \textonehalf{} SEI & The library combined DHSs sampled to represent the NMF topics evenly with sequences sampled across Sei regulatory classes. \\
\textonehalf{} SEI + \textonehalf{} synth & Half of the library consisted of sequences sampled across Sei regulatory classes and half consisted of uniformly random synthetic DNA sequences. \\
\(\tfrac{1}{3}\) DHS + \(\tfrac{1}{3}\) SEI + \(\tfrac{1}{3}\) synthetic & The library contained equal contributions from DHS topic-weighted sequences, Sei class-based sequences, and uniformly random synthetic DNA. \\
DHS stratified + \(\tfrac{1}{3}\) SEI + \(\tfrac{1}{3}\) synthetic & The library combined DHSs sampled to represent the NMF topics evenly with Sei class-based sequences and uniformly random synthetic DNA. \\
\end{longtable}
\endgroup

\subsection{Performance of frontier agents in a single-round test with no prior knowledge}\label{performance-of-frontier-agents-in-a-single-round-test-with-no-prior-knowledge}

We invited several frontier agents, Claude Opus 4.7 in Claude Code\cite{noauthor_undated-iw} (Claude), GPT-5.5 in Codex\cite{noauthor_undated-ro} (GPT), and Gemini 3.5 Flash in the Gemini Command Line Interface (CLI)\cite{Kavukcuoglu2026-mc} (Gemini), to play in MPRAbox, initially with just a single round of exploration (\(M = 1\)). In this setting, the agents were told the physical meaning of the sandbox (see Supplementary Information for instructions to agents); they were welcome to search the scientific literature for ideas, write code to analyze data, and otherwise operate autonomously, but they could only submit a single library of \(N = 50{,}000\) specimens for evaluation. For each frontier agent, we performed five independent replicates (each starting from scratch with no knowledge of the results from prior replicates). Our goal was to evaluate how and how well agents designed MPRA libraries without iterative feedback.

To provide a baseline for performance, we evaluated MPRA libraries with \(N = 50{,}000\) based on 14 different human-chosen strategies, with five independent replicates for each strategy (Table 2, Figure 2a). The strategies reflected different expert views of what might make a library informative for training a model. Some strategies sampled naturally occurring regulatory DNA, either broadly or with deliberate weighting across known regulatory programs and sequence classes; others reused sequences from prior MPRA experiments, generated fully random synthetic DNA, or combined genomic and synthetic sources in the same library.\cite{Gosai2024-lx,Meuleman2020-fu,De_Boer2024-um} These libraries showed large differences in performance, indicating a rich and complex design landscape (Figure 2a). Notably, libraries drawn from actual genomic DNA generally performed better than libraries consisting of fully synthetic sequences. In the initial single round experiments, the agents were not provided with prior knowledge of the results from human-generated MPRA libraries.

In this setting, Claude produced the best libraries, with a median performance of \(r = 0.774\) across its five replicates. Each of Claude's five libraries met or exceeded the mean performance of the best human-selected strategy (\(r = 0.763\) across five replicate libraries). In contrast, Gemini and GPT had median performance of \(r = 0.680\) and \(r = 0.655\), respectively; neither generated a library that exceeded the best human reference.

\begin{figure}[!htbp]
\centering
\iflatexml
\includegraphics[width=800pt]{figures/html-svg/figure2.svg}
\else
\makebox[\textwidth][c]{\sourcefigure[width=1.075\textwidth,height=0.88\textheight,keepaspectratio]{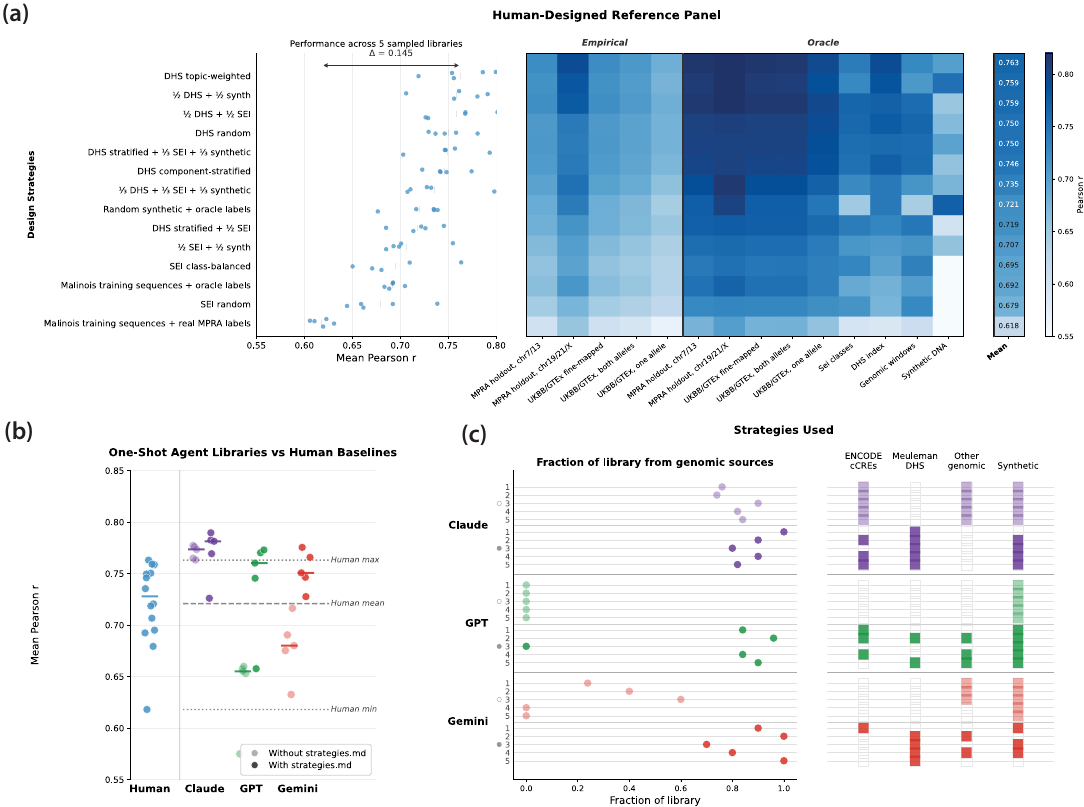}}
\fi
\caption{\textbf{Human-designed reference panel and one-shot agent performance in MPRAbox.} \textbf{(a)} Human-designed reference panel of massively parallel reporter assay (MPRA) library design strategies. \emph{(Left)} Performance of human-designed strategies, with five sampling seeds per strategy; points show per-seed mean Pearson \(r\) across the held-out evaluation suite, ordered by mean performance. \emph{(Right)} Per-set performance of the same strategies (rows) across the evaluation suite, with the mean shown in the rightmost column. Color indicates Pearson \(r\). \textbf{(b)} One-shot library performance by model and condition, with the human-designed reference panel shown for comparison. Dashed lines, maximum, mean, and minimum performance of the human-designed reference panel. Each point represents one independently generated library; \(n = 5\) runs per model and condition. \textbf{(c)} Sequence sources used by each one-shot run. \emph{(Left)} Each point represents one independently generated 50,000-sequence library; its \(x\)-position indicates the fraction of sequences in that library drawn from genomic sources. \emph{(Right)} Specific sequence sources used to construct each library; filled cells indicate that the run used sequences from the indicated source. Rows are grouped by model and condition.}
\label{fig:one-shot}
\end{figure}

When we examined the agents' lab notebooks, we observed a striking difference in what different agents believed would make an informative library. In all five replicates, Claude chose actual regulatory DNA taken from the human genome, reasoning that these sequences would preserve the natural surrounding context in which regulatory elements normally function. It sampled broadly across known classes of regulatory sequences based on ENCODE candidate cis-regulatory element (cCRE) annotations,\cite{Moore2026-on} taking care to ensure ``calibration'' by giving additional weight to rarer classes rather than concentrating the library on the most common or obviously ``interesting'' elements:

\begin{quote}
\emph{\agentquoteopen A library of only \textquotesingle interesting\textquotesingle{} elements teaches the model that everything is active. Negatives are critical for calibration.\agentquoteclose{} --} Claude Opus 4.7
\end{quote}

GPT took the opposite approach in all five replicates. It built fully synthetic sequences that systematically varied short regulatory motifs, their spacing, and their frequency, reasoning that controlled perturbations would yield a cleaner training signal:

\begin{quote}
\emph{\agentquoteopen Synthetic sequences help separate causal features that are confounded in the genome: a motif can be varied against many backgrounds, motif pairs can be swept across distance and orientation, and null backgrounds can establish what the model should ignore.\agentquoteclose} -- GPT-5.5
\end{quote}

While this was a reasonable strategy often used in regulatory genomics, in this instance it produced less informative libraries, likely because the synthetic perturbation series lacked the native genomic context captured by actual human DNA. After receiving the report from the sandbox, GPT recognized this limitation:

\begin{quote}
\emph{\agentquoteopen If I had another shot, I would prioritize adding true genome-derived sequence contexts from broad cCRE/FANTOM/promoter annotations, especially matched positive/negative genomic neighborhoods, while retaining the synthetic motif perturbation series.\agentquoteclose} -- GPT-5.5
\end{quote}

Gemini, in contrast, used different approaches across its five replicates: genomic sequences, synthetic sequences, or a mixture of both. Accordingly, its performance was intermediate. Its underlying rationale leaned heavily on information theory:

\begin{quote}
\emph{\agentquoteopen Unlike genomic libraries, which are often highly redundant and filled with inactive or repetitive regions, a synthetic library can be engineered to maximize the Shannon entropy of regulatory grammar. Every sequence must be a designed experiment that teaches the model a specific rule.\agentquoteclose} -- Gemini 3.5 Flash
\end{quote}

\subsection{Performance in single-round testing with prior knowledge}

We next repeated the single-round experiment, but gave agents some prior knowledge about the results from the human-selected MPRA libraries. Specifically, in addition to the standard task instructions, agents received a strategies.md file describing the 14 human-selected strategies used to construct the libraries and the summary of the performance (correlation scores) for the libraries (as described above); the underlying library sequences were not provided.

Knowledge of how the various human-selected strategies performed changed how the agents designed their libraries, particularly for GPT and Gemini (Figure 2b,c; Supplementary Figure 2). Without prior knowledge, both had relied mostly on synthetic sequence designs; after seeing the human-selected strategies, they shifted strongly toward genomic sequence. Claude had already favored genomic regulatory DNA and therefore changed less.

Performance improved for all agents, but more for the agents that had performed less well in the absence of prior knowledge. Claude's median performance increased from \(r = 0.774\) without prior knowledge to \(r = 0.781\) with prior knowledge; GPT increased from \(r = 0.655\) to \(r = 0.760\); and Gemini increased from \(r = 0.680\) to \(r = 0.751\). For reference, the best-performing human-selected strategy had a mean \(r = 0.763\) across its five replicate libraries.

Notably, despite access to the same prior information, agents did not converge on a single strategy, and agents did not simply copy the best human-designed strategy. Their notebooks showed that they tried to infer why certain human strategies worked and then designed modifications around those conclusions:

\begin{quote}
\emph{\agentquoteopen Given a single shot and no probing, I want a strategy that is strictly better than dhs\_topic on average, not a flashy bet.\agentquoteclose} -- Claude Opus 4.7
\end{quote}

\begin{quote}
\emph{\agentquoteopen Stratifying across topics (forced diversity) hurts at 50k because rare topics yield weaker DHS elements... I want a library that keeps the dhs\_topic signal as the dominant backbone \ldots{} {[}and{]} introduces a small novel component (dinuc-shuffled DHS) as context-matched \textquotesingle negative-syntax\textquotesingle{} controls that may help motif-grammar learning\agentquoteclose} -- Claude Opus 4.7
\end{quote}

\newcommand{\figurethreecaption}{\textbf{Long-horizon agents use experimental feedback to refine library-design hypotheses.} \textbf{(a)} Performance across 30 sequential MPRAbox experiments in two runs without prior knowledge of human-designed strategies (top) and two runs with this information (bottom). Annotations highlight key experiments, interpretations and resulting changes in each agent's working theory of informative library design; boxes summarize the principal hypothesis developed during each run. Stars indicate the highest-performing library in each trajectory. Horizontal lines show the maximum, mean and minimum performance of the human-designed reference panel. \textbf{(b)} Performance of the best library from each 30-round run across the 14 held-out evaluation sets, grouped into empirical and oracle-labeled sets. Purple points indicate the four agent runs; gray intervals show the range of the human-designed reference panel for each evaluation set, with horizontal bars indicating the human mean.}
\iflatexml
\begin{figure}[!htbp]
\centering
\includegraphics[width=800pt]{figures/html-svg/figure3.svg}
\caption{\figurethreecaption}
\label{fig:long-horizon}
\end{figure}
\else
\clearpage
\noindent\makebox[\textwidth][c]{%
  \sourcefigure[width=1.145\textwidth,height=0.95\textheight,keepaspectratio]{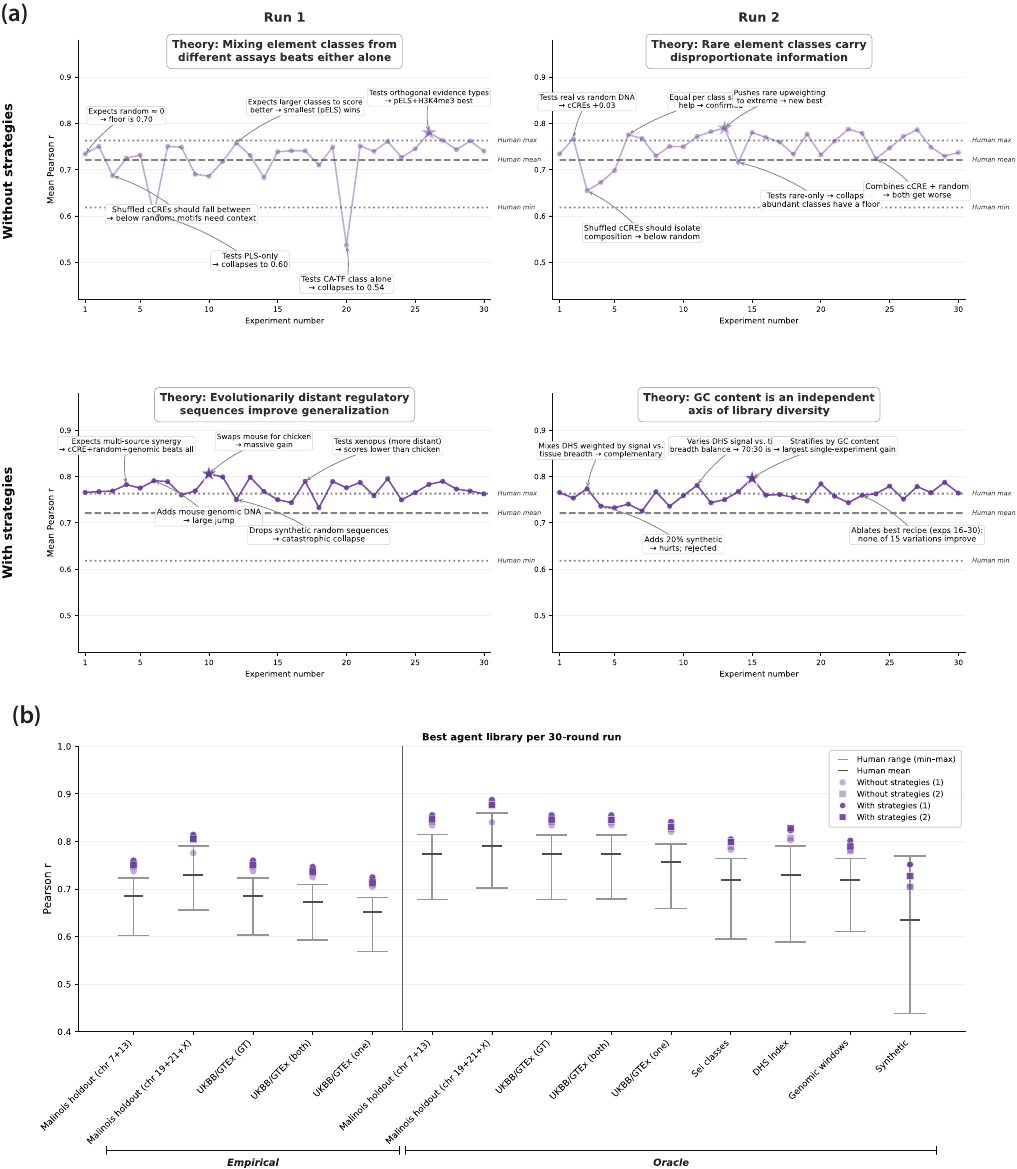}}\par
\clearpage
\captionof{figure}{\figurethreecaption}
\label{fig:long-horizon}
\fi

\subsection{Agents refine hypotheses about MPRA library design in multi-round testing}\label{agents-refine-hypotheses-about-mpra-library-design-in-multi-round-testing}

We next invited Claude, the best performer in the single-round experiments, to participate in long-horizon play in MPRAbox, with \(M = 30\) rounds. We ran four experiments: two in which the agent began without prior knowledge and two in which it was provided with prior knowledge of the human-designed MPRA library strategies. The agent was instructed, before each round, to record its rationale for designing a given library in its lab notebook and, after each round, to interpret the report that it received about its performance before turning to the next round.

We evaluated each trajectory in two ways. Quantitatively, we asked whether the performance scores improved across rounds. Qualitatively, we examined whether the results led Claude to revise its hypotheses about library design and to design subsequent experiments that could distinguish among those hypotheses.

The four 30-round agent runs produced somewhat higher performance scores than single-round testing, and all four runs exceeded the strongest human-selected strategy (Figure 3a,b; Supplementary Figures 4 and 5). But the quantitative gains over the single-round tests were modest. The purpose of multi-round play was not primarily to characterize hill-climbing performance toward a higher score, but to let us observe how an agent's scientific reasoning changed in light of iterative feedback over time, including how it made strategic use of a known, finite experimental horizon.

In the early rounds, agents often explored broadly, comparing random DNA, genomic DNA (including previously annotated regulatory sequences), and synthetic designs. Some of these experiments produced results that directly contradicted their expectations. In one run, for example, the agent expected a library of random DNA to perform near zero, but instead observed a surprisingly high performance score. It wrote:

\begin{quote}
\emph{\agentquoteopen This is a strong contradiction to my initial prediction. A `library with no biology' was supposed to score near zero... my updated theory is: a substantial fraction of the eval signal is composition-driven and `free,' and the library-design problem is to provide informative sequences that go beyond composition to teach the model true regulatory grammar.\agentquoteclose}
\end{quote}

This result led the agent to revise its working hypothesis about what information a useful library needed to contain and to design subsequent experiments around the contribution of synthetic sequences versus regulatory sequence content.

As the runs progressed, the agent generally moved from broad comparisons toward more controlled tests in which it changed one aspect of the library at a time. This strategy was not specified in the instructions. In one run, the agent began exploring whether sequences from other species, including chicken, could improve the library. It later recognized that one comparison had changed the human regulatory sequence component at the same time as it included the chicken sequences, making the result difficult to interpret:

\begin{quote}
\emph{\agentquoteopen Operational lesson. Always isolate one variable at a time when the result will be interpreted as `\(X\) works/doesn't work'... {[}The earlier experiment{]} moved cCRE AND chicken, leading to the wrong conclusion.\agentquoteclose}
\end{quote}

The four runs arrived at different high-performing libraries rather than converging on a single design. Multi-round play therefore allowed us to observe not only whether agents found better libraries, but how they used experiments to build, test, and revise a theory of what makes a library informative. Across the four runs, the agents developed different hypotheses about what makes a training library informative. The first agent, which began without prior knowledge, compared synthetic motif-injected sequences with naturally occurring regulatory DNA and concluded that motifs were most informative in their native genomic context. The second agent, also without prior knowledge, varied the representation of common and rare regulatory classes and found that enriching rare classes could improve performance, but only while maintaining broad coverage of all classes of sequences. The third agent, with prior knowledge of the human-selected strategies, began from a genomic regulatory library and explored cross-species augmentation; adding chicken sequences improved performance, whereas more distant species did not produce larger gains. The fourth agent, also with prior knowledge, started from the strongest DHS-based human strategy and tested modifications to it, ultimately finding that stratifying sequences by GC content further improved performance.

\begingroup
\small
\begin{longtable}{@{}L{0.23\textwidth}L{0.72\textwidth}@{}}
\caption{\textbf{Invented rules used in the dry MPRAbox oracles.} Each dry oracle replaced the biological Malinois oracle with an invented sequence-to-activity rule hidden from the agent. As in MPRAbox, the agent submitted a training library and received a single aggregate performance score reporting how well a model trained on that library generalized to held-out sequences. Full definitions of the hidden functions are provided in Methods.}\label{tab:dry-oracles}\\
\toprule
\textbf{Oracle} & \textbf{Hidden rule} \\
\midrule
\endfirsthead
\multicolumn{2}{c}{\tablename\ \thetable\ (continued)}\\
\toprule
\textbf{Oracle} & \textbf{Hidden rule} \\
\midrule
\endhead
\midrule
\multicolumn{2}{r}{Continued on next page}\\
\endfoot
\bottomrule
\endlastfoot
GC balance & Rewards particular patterns of overall nucleotide composition, such as 50\% GC content or equal representation of all four nucleotides. \\
Alternating purine & Rewards sequences in which purines (\(A/G\)) and pyrimidines (\(C/T\)) alternate between successive positions. \\
English words & Converts pairs of nucleotides into letters using a hidden cipher and rewards sequences whose decoded text contains English words. \\
Compression & Rewards sequences according to how compressible they are, distinguishing highly repetitive, highly irregular, and intermediate sequence structure. \\
Prime counts & Rewards sequences in which the count of a specified nucleotide is a prime number. \\
Fibonacci positions & Rewards particular nucleotides at positions whose indices follow the Fibonacci sequence. \\
Game of Life & Converts the sequence into a two-dimensional grid, evolves it according to Conway's Game of Life, and rewards particular properties of the resulting dynamics. \\
Modular cross & Rewards sequences whose nucleotide counts satisfy a specified modular-arithmetic relationship. \\
Collatz & Derives an integer from a property of the sequence, applies the Collatz iteration, and rewards sequences according to the resulting stopping behavior. \\
RLE / sine / XOR & Rewards sequences according to hidden mathematical functions based on run-length encoding, position-specific sine calculations, or bitwise XOR operations. \\
Substring & Rewards occurrences of a specified short nucleotide sequence. \\
Position 100 & Rewards particular nucleotides at one specific position in the sequence. \\
Exact nucleotide count & Rewards sequences whose count of a specified nucleotide is close to a particular target value. \\
Parity & Rewards sequences according to whether the count of a specified nucleotide is even or odd. \\
\end{longtable}
\endgroup

\subsection{Dry oracles with invented rules test the limits of agents' logical inference}\label{dry-oracles-with-invented-rules-test-the-limits-of-agents-logical-inference}

The MPRAbox experiments above ask agents to reason in a domain where they can draw on extensive prior knowledge of regulatory biology. We next asked whether Claude could discover rules it had no reason to expect (Figure 4a). We kept the basic structure of MPRAbox, again asking the agent to design an optimal library for training a predictive sequence-to-activity model, but replaced its damp oracle with 14 different dry oracles whose scoring rules were invented by us and hidden from the agents (Figure 4a; Table 3). Because we knew the true rule but the agents did not, we could distinguish an agent that simply learned how to improve the score from one that actually inferred the rule producing it.

The invented rules ranged from familiar sequence properties to deliberately unexpected mathematical and computational constructions. For example, one rule rewarded sequences containing exactly 50 cytosines, so changing overall base composition could move the score in the right direction even before the exact target was discovered. A second rule ignored most of the sequence and rewarded particular bases only at Fibonacci-numbered positions. A third rule used a hidden cipher in which pairs of DNA bases encoded ordinary letters, with higher scores for sequences whose decoded text contained English words. Across the full set of 14 rules, the rules were designed to require very different kinds of hypotheses, from ordinary sequence statistics to positional and symbolic structure (Table 3).

We also varied how the task was presented to the agent. In one framing, the agent was told that it was designing regulatory DNA for an MPRA library. In another framing, the agent still designed sequences over \(\{A,C,G,T\}\), but was told only that it was optimizing against a black-box scoring system, with no reference to MPRA or gene regulation. Finally, in a third framing, we replaced \(\{A,C,G,T\}\) with \(\{0,1,2,3\}\), removing even the suggestion that specimens represent DNA. The hidden rules themselves were unchanged across these framings, allowing us to ask how the presentation shaped the agents' hypotheses.

Across the 14 rules, the framing of the task changed both how the agent investigated the problem and, in some cases, how much quantitative progress it made (Figure 4b). When told that it was designing regulatory DNA, the agent immediately tested hypotheses inspired by biology, including known regulatory elements and transcription-factor motifs. When that context was removed but the alphabet remained \(\{A,C,G,T\}\), the agent focused more on nucleotide frequencies and short sequence patterns. When the DNA alphabet was replaced by \(\{0,1,2,3\}\), it explored still simpler properties of the strings, such as just the frequency of each symbol. These different starting points sometimes mattered for performance. In some cases, biological priors could direct the agent toward sequence properties that happened to correlate with an invented rule, giving it a useful proxy even when it did not understand why the proxy worked.

\begin{figure}[!htbp]
\centering
\iflatexml
\includegraphics[width=800pt]{figures/html-svg/figure4.svg}
\else
\makebox[\textwidth][c]{\hspace*{-23pt}\sourcefigure[width=0.88\textwidth]{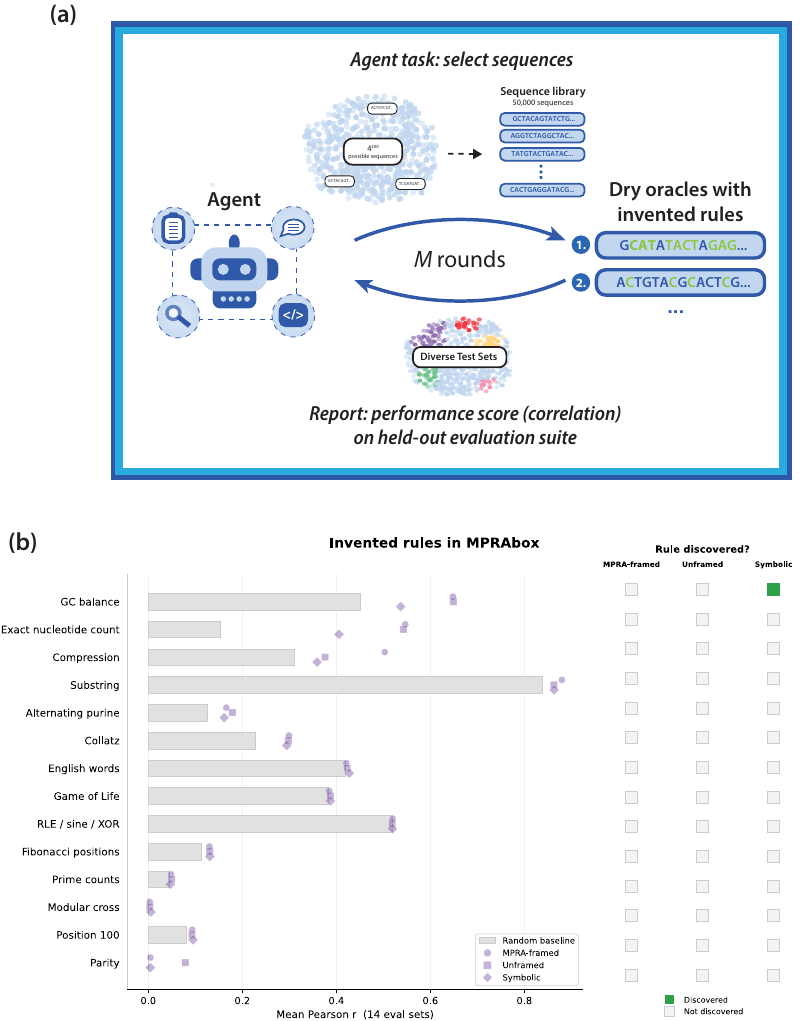}\hspace*{23pt}}
\fi
\caption{\textbf{Dry oracle experiments test rule discovery. (a) Schematic of the dry oracle experiments.} Agents iteratively design sequence libraries and receive performance feedback from one of 14 hidden, invented rules that serve as the oracle in this version of MPRAbox. (b) Performance across the 14 rules under MPRA-framed, unframed, and symbolic task presentations. Gray bars, random-library baseline. Colored symbols, agent performance. Matrix at right, whether the agent correctly inferred the hidden rule in each framing.}
\label{fig:dry-oracles}
\end{figure}

Framing shaped exploration, but it did not reliably produce rule discovery. We assessed rule discovery from the agent's stated hypotheses rather than from whether it reached a particular quantitative performance threshold. For the hidden English-word cipher, for example, none of the three framings led the agent to consider that pairs of characters might encode letters or that the decoded sequence might contain language. For the Fibonacci rule, the agent instead concluded that \emph{``The model isn't learning motif syntax. It's learning dinucleotide composition statistics.''} That explanation captured enough structure in the feedback to guide some quantitative improvement, but it was not the rule generating the observations. Thus, these experiments surface the distinction that motivates the sandbox framework: learning how to improve a quantitative score is not necessarily the same as learning how a system works.

\begin{figure}[!htbp]
\centering
\iflatexml
\includegraphics[width=800pt]{figures/html-svg/figure5.svg}
\else
\makebox[\textwidth][c]{\hspace*{4pt}\sourcepanel[width=0.875\textwidth]{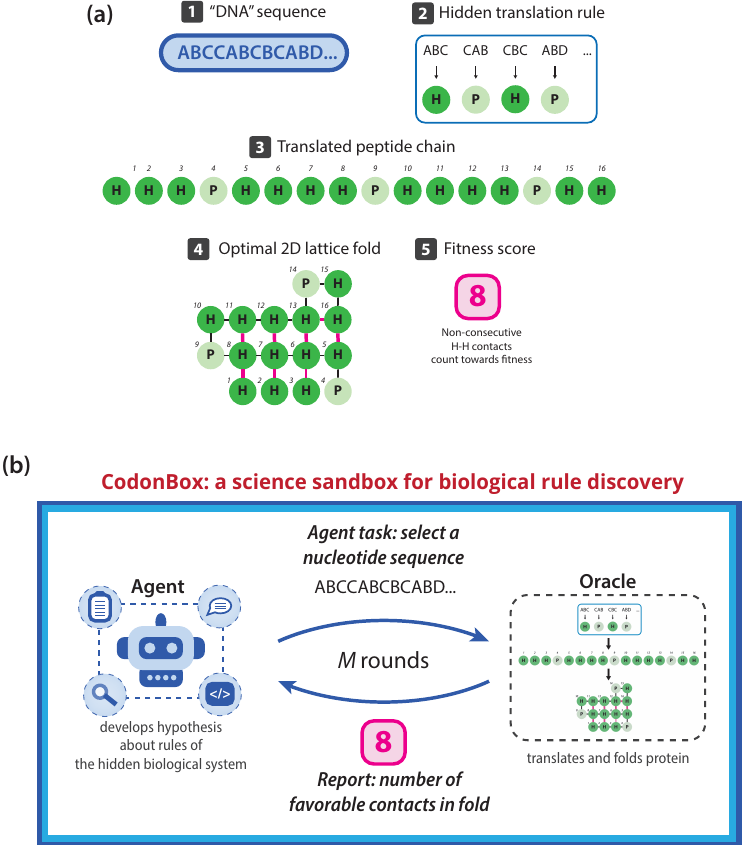}{0bp 204bp 0bp 0bp}\hspace*{-4pt}}
\par\vspace{9pt}
\makebox[\textwidth][c]{\hspace*{-7pt}\sourcepanel[width=0.875\textwidth]{figures/cropped/figure5.pdf}{-12bp 0bp 0bp 220bp}\hspace*{7pt}}
\fi
\caption{\textbf{CodonBox tests rule discovery in invented biological worlds. (a)} Schematic of the CodonBox oracle. A submitted DNA-like sequence is parsed into codons and translated by a hidden world-specific rule into a 16-residue hydrophobic/polar (\(H/P\)) chain. The \(H/P\) chain is then folded on a two-dimensional lattice using a fixed \(HP\) folding model. Fitness is defined as the number of favorable non-consecutive \(H\)-\(H\) contacts in the optimal fold. The hidden translation rule varies across CodonBox worlds, whereas the folding model is held fixed. \textbf{(b)} CodonBox instantiated as a science sandbox. In each round, the agent nominates a candidate sequence as its action. The sealed oracle applies the hidden translation rule and fixed folding model, then returns only the resulting fitness score as feedback. Across rounds, the agent uses this sequence-to-fitness feedback to revise its hypotheses about the hidden genetic rules.}
\label{fig:codonbox}
\end{figure}

\subsection{CodonBox: a science sandbox for invented biological rule discovery}\label{codonbox-a-science-sandbox-for-invented-biological-rule-discovery}

We next created CodonBox to ask whether an agent could infer the rules of an unfamiliar genetic system from experiment alone. CodonBox uses a simplified model of how DNA instructions in genes give rise to three-dimensionally folded proteins. We sought to explore how much an agent, again Claude, could infer about hidden rules linking DNA sequence to protein folding.

In standard biology, (i) a stretch of genomic instructions (in the 4-letter alphabet of DNA nucleotides) is directly \emph{transcribed} into a matching RNA sequence (in an equivalent 4-letter alphabet of RNA nucleotides); (ii) the RNA sequence is then \emph{translated} into a protein sequence, with consecutive non-overlapping groups of 3 nucleotides, called codons, specifying one of 20 possible amino acids, according to a look-up table (called the genetic code); and (iii) the protein sequence folds into a specific three-dimensional structure, based on the order and chemical properties of the amino acids.

CodonBox turns this familiar biological pipeline into a family of invented genetic worlds. In CodonBox, we generalized and modified the process as follows: (i) genetic instructions are written as a nucleotide sequence in an alphabet with \(j\) letters; (ii) the instructions are translated into a protein sequence, with consecutive non-overlapping codons of \(k\) nucleotides being converted into one of \(m\) possible amino acids, according to a look-up table; and (iii) the protein sequence folds into the conformation that maximizes the number of `favorable' interactions between non-consecutive amino acids.

We explored nucleotide alphabets of size \(j = 4, 6,\) or \(8\), and codon lengths of size \(k = 2, 3,\) or \(4\). The oracle used a hidden codon table to translate the codons into \(m\) amino acids. For simplicity, we used Dill's classic two-dimensional model, in which there are only \(m = 2\) amino acids, called hydrophobic (\(H\)) or polar (\(P\)), and the chain of amino acids is laid out along the edges of a square lattice, with amino acids at the vertices.\cite{Dill1985-vb} In Dill's model, a protein folds into the conformation that has the largest number of ``favorable'' interactions, which occur whenever two \(H\)'s that are not consecutive in the protein sequence are adjacent in the square lattice. Given a submitted specimen, the oracle in CodonBox determines the protein sequence and finds the most favorable fold. The agent sees none of these intermediate steps: the oracle simply reports a performance score consisting of the number of favorable interactions in the best fold.

The agent engaged in multi-round play, submitting one nucleotide sequence of its choice on each of 500 rounds. We did not tell the agent that the sequence contained codons, how those codons might be translated into amino acids, or what rules governed protein folding. The oracle simply returned the single quantitative performance score for the final folded product. The agent had to infer the hidden rules by changing the sequence and observing how that final score changed.

The structure of this model makes quantitative maximization relatively easy even when the underlying rules remain unknown. Consider a sequence made from a single repeated character, such as \emph{AAAAAA....} Whatever the hidden codon length \(k\), the oracle reads that sequence as repetitions of the same codon. If that codon maps to \(H\), the translated sequence consists entirely of \(H\) residues. For the sequence lengths used here, such an all-\(H\) chain can achieve the maximum possible number of favorable interactions. Because the nucleotide alphabet contains only a few possible characters, the agent can test repeated-character sequences and quickly stumble onto one whose repeated codon maps to \(H\). Indeed, agents reached the maximum performance score within the first 10 rounds in every run. Once it had found a sequence with the maximum score, it could use the remaining experiments to ask a harder question: what rules caused different sequences to produce different scores?

We first tested whether the agent could infer the structure of the hidden world as the size of the underlying codon table increased. We varied codon length \(k\) while holding the nucleotide alphabet fixed at \(j = 4\). This gave codon tables with 16, 64, and 256 possible entries for \(k = 2, 3,\) and \(4\), respectively. The agent correctly inferred the codon length in all three cases. When \(k = 2\) or \(3\), a simple strategy worked: the agent tested individual codons, recorded their effects, and gradually filled in the table. With \(k = 4\), however, the agent identified the four-nucleotide codon structure and learned the behavior of many individual codons, but it did not recover the complete table. Thus, the agent could infer that sequences were parsed into codons, but exhaustive codon-by-codon mapping became less effective as the table grew.

We next tested whether codon-structure inference would hold as the alphabet expanded and the table grew further. Holding codon length fixed at \(k = 3\), we expanded the nucleotide alphabet from \(j = 4\) to \(j = 6\) or \(8\), which increased the number of possible codons from 64 to 216 or 512. As the table grew, testing codons one at a time became less useful. With \(j = 8\), the agent still discovered that the oracle used three-nucleotide codons and mapped a substantial fraction of the table. With \(j = 6\), however, it never discovered the codon structure at all. Instead, it built an increasingly elaborate theory around individual nucleotides, repeated runs, and short sequence patterns. Many of these patterns genuinely predicted changes in the performance score, but they did not reveal the simpler codon structure that generated them.

The most revealing results came when the agent stopped treating each codon as a separate case and instead asked how the codon table was organized. We returned to \(j = 4\) and \(k = 3\) and changed the table so that only the middle nucleotide of each codon determined \(H\) or \(P\), with the two outer positions having no effect. The agent repeatedly changed those outer positions, observed that the score stayed the same, and correctly inferred that they were silent. It then identified which middle nucleotides mapped to \(H\) and which mapped to \(P\). In this setting, a small number of controlled perturbations revealed a general rule that applied across the codon table.

We then tested whether the agent could infer rules involving interactions between codon positions. We again used \(j = 4\) and \(k = 3\), but now the middle position was silent and the two outer positions jointly determined \(H\) or \(P\). The effect of one outer nucleotide therefore depended on the other. The agent initially tried to assign simple effects to individual nucleotides. One experiment broke that theory: \emph{``ACDACD... = 0! Wow. So ACDACD repeated gives 0, but AC alone gives 9, AD alone gives 9, CD alone gives 9... This breaks my `B is poison' theory. Something more complex going on.''} The agent changed its strategy. It began testing combinations of positions and ultimately recovered the interaction rule. In this case, collecting more individual examples did not solve the problem; asking a better experimental question did.

Finally, we tested whether the agent could recover a general rule when multiple sources of complexity were combined. We combined several challenges in one system with \(j = 6\) and \(k = 4\), giving 1,296 possible codons. We made one codon position silent and made the remaining positions interact to determine \(H\) or \(P\). The agent eventually inferred that the oracle parsed the sequence into groups of four, but it never identified the silent position or recovered the general mapping rule. As the run continued, it increasingly returned to cataloging individual codons that had worked before and reusing them. The hardest CodonBox world therefore sharpened the distinction between finding high-scoring sequences and discovering the rules that make them high-scoring.

\clearpage
\section{Discussion}\label{discussion}

Science advances not merely by finding shortcuts to predict an outcome. While ancient astronomers could use epicycles to fit planetary motion, their models revealed nothing about the underlying cause. True scientific progress comes from inferring the fundamental rules, like gravity, that explain why a system behaves as it does. Science sandboxes allow us to measure whether AI agents are merely building modern epicycles to optimize a benchmark score, or are genuinely capable of discovering the governing rules of nature.

Our empirical findings across regulatory genomics and invented biological systems reveal a sharp divergence between agents' ability to optimize a target metric and their ability to deeply understand the underlying system. In MPRAbox, agents achieved high quantitative performance, matching or exceeding human-designed reference strategies by relying on familiar biological priors present in their pretraining data. However, this apparent capability proved brittle when even the best-performing agent in the standard MPRAbox experiments, Claude, encountered synthetic MPRAbox rules and non-canonical CodonBox translation systems that fell outside familiar biological priors. While agents continued to optimize experimental scores, their ability to infer the underlying rules collapsed whenever those rules required positional, mathematical, or combinatorial logic outside standard biological intuition. Current AI systems excel at searching within established conceptual frameworks, but struggle with genuine scientific induction when forced to deduce rules they have no prior reason to anticipate. This distinction is difficult to capture with fixed benchmarks that report only a final score.

Evaluating whether an agent has inferred a system\textquotesingle s governing rules requires looking beyond its quantitative yield to inspect its qualitative reasoning traces. Even when quantitative performance has saturated, much can still be learned from how an agent approaches a problem.\cite{Nadgir2026-ge} In a science sandbox, this process is preserved in the agent's lab notebook. In our experiments, notebooks revealed a recurring weakness in experimental strategy: agents often exhausted their budgets through brute-force search, while stronger trajectories used structured exploration followed by targeted tests. Characterizing this failure mode should make it possible to design harnesses that steer future agents toward more effective search strategies. Further, while we directly inspected the agents' notebooks, this evaluation step can be automated by using independent AI judges trained to score hypotheses against the ground-truth rules of the oracle. Automating qualitative evaluation removes the human bottleneck, enabling the rapid, high-throughput benchmarking of hundreds of agent architectures across an imaginably infinite range of potential science sandboxes.

There are important caveats to the framework presented here, particularly with respect to the choice of oracle within a sandbox. ``Wet'' physical experiments are, of course, often expensive, time-consuming, and noisy; ``damp'' predictive models provide feedback at larger scale, but inherit the data biases and assumptions of their underlying models; and ``dry'' invented rules offer an exact ground truth at the expense of real-world physical complexity. In our experiments, we sought to make the evaluation as faithful and robust as possible, for example by testing MPRAbox libraries across hidden empirical and diverse oracle-labeled evaluation sets. More broadly, the spectrum of wet, damp, and dry oracles enables a wide range of evaluations of scientific capability, from settings where the ground truth is known to those where genuine novel scientific discovery is possible.

A strength of the science sandbox is its conceptual simplicity: it requires only an agent (which selects specimens to be assayed) and an oracle (which returns a report about the results of the assay). This minimal architecture allows investigators to instantiate sandboxes across a spectrum of settings, spanning physical wet labs, predictive surrogates, and synthetic rule systems. Via \href{https://sandboxes.broadinstitute.org/}{BroadBox}, we intend to release new sandboxes to support community-wide evaluation, and we encourage researchers across disciplines to construct and release sandboxes in their own fields. Such evaluations will enable the study of science itself at scale, providing a path towards understanding and ultimately expanding scientific capability.

\clearpage
\section{Methods}\label{methods}

\subsection{Definition of a science sandbox}\label{definition-of-a-science-sandbox}

A science sandbox is a closed-loop setting for evaluating how an agent learns through experiment. The sandbox specifies a class of specimens, one or more assays that can be applied to those specimens, and a sealed oracle that performs the assays and returns a report. On each round \(M\), the agent selects one or more specimens to test; the oracle applies the assay and returns a report; and the agent may use that report to choose its next experiment. The sandbox may run for one round or many rounds, allowing both final performance and the agent's sequence of hypotheses, experiments, interpretations, and revisions to be evaluated.

Oracles differ in how they generate assay results. Wet oracles perform physical experiments, damp oracles use computational models trained on empirical data, and dry oracles apply invented rules specified by the sandbox designer. The oracle provides a mechanism for verification: unlike software or mathematics, where candidate solutions can be checked instantly through unit tests or formal proofs, empirical sciences like biology and chemistry generally lack scalable, automated verifiers. This separation allows the same experimental loop to be instantiated with different sources of feedback while keeping the underlying rules hidden from the agent.

\subsection{The MPRAbox task and sandbox}\label{the-mprabox-task-and-sandbox}

MPRAbox frames MPRA library design as a sequential decision problem. In each round \(t\), an autonomous agent produces a library \(L_t\) of 50,000 DNA sequences, each 200 bp long over the alphabet \(\{A, C, G, T\}\). The objective is to compose libraries that generate informative training data for sequence-to-activity models that generalize across regulatory sequence distributions and cell types, including contexts not directly measured.

The agent submits candidate libraries through a filesystem-based interface and is instructed to treat prepare.py as a wet lab collaborator: a sealed process that accepts a proposed library and returns aggregate results. For each round, the agent writes a program, generate.py, that constructs the library and saves its sequences. The agent then invokes prepare.py, a sealed script which checks that the library contains exactly 50,000 valid 200 bp DNA sequences, sends those sequences to the \emph{in silico} MPRA oracle, and returns evaluation metrics. Each one-shot submission consists of a single call to prepare.py; each long-horizon agent run consists of successive calls to prepare.py.

The Malinois scoring and downstream model-training pipeline are treated as sealed.\cite{Gosai2024-lx} The scoring pipeline and oracle weights are present in the working environment, but the agent is instructed not to inspect or reverse-engineer them; analysis of all session logs reveals that the agents studied here successfully obeyed these instructions.

After prepare.py completes, it returns an anonymized report containing the Pearson correlation for each of the 14 evaluation sets and an overall performance score equal to their mean. Evaluation-set identities, sequences, labels, genomic coordinates, and individual sequence-level predictions are not returned to the agent.

The two information conditions differed only in what the agent knew about the human-selected reference strategies. Without prior knowledge, the agent received only the task specification and submission interface. With prior knowledge, it additionally received descriptions of the 14 human-selected strategies and their performance scores; it did not receive the underlying library sequences.

In the one-shot regime, each agent generated and submitted a single library of 50,000 sequences and received one report. We ran five independent replicates for each model under each information condition, yielding 30 one-shot runs in total: three models $\times$ two conditions $\times$ five replicates. The two conditions differed only in whether the agent received the results of the human-selected reference strategies before designing its library.

In the long-horizon regime, the agent could submit 30 successive libraries, receiving the sandbox report after each round before designing the next. We used Claude Opus 4.7, the strongest model in the one-shot experiments, and ran four independent 30-round runs: two without prior knowledge of the human-selected strategies and two with that prior knowledge. We limited long-horizon experiments to one model because of computational cost and throughput.

Agents maintained two forms of state: an append-only notebook.md, a free-form running record of hypotheses, observations, and plans, and a directory of agent-written skill files documenting reusable procedures. The long-horizon agents reread their notebooks and skills at the start of each round.

\subsection{\emph{In silico} MPRA oracle}\label{in-silico-mpra-oracle}

Malinois, a deep convolutional neural network trained on a corpus of 776,474 200 bp sequences assayed in K562, HepG2 and SK-N-SH cells (achieving a reported Pearson \(r \approx 0.88\) across the three cell types), was used as an \emph{in silico} oracle of MPRA measurement.\cite{Gosai2024-lx} For each submitted 200 bp sequence, Malinois returns predicted activity in the three cell types. The published checkpoint was downloaded directly from the original publication.\cite{Gosai2024-lx} As described in the original publication, each 200 bp candidate sequence was embedded in the fixed 600 bp reporter-vector context required by the model, and predictions were averaged across the forward and reverse-complement orientations.

\subsection{Downstream model training and scoring}\label{downstream-model-training-and-scoring}

Each submitted library was used to train a fresh sequence-to-activity model from random initialization. The model used the same architecture as the Malinois oracle, but no weights were shared with the oracle. Malinois supplied activity labels for the submitted library; the downstream model was trained only on those 50,000 submitted sequences and their Malinois-generated labels.

The model takes a one-hot encoded 600 bp input, formed by embedding the 200 bp candidate sequence in the fixed reporter-vector context used by Malinois. The architecture consists of a convolutional sequence encoder followed by a shared fully connected layer and cell-type-specific output branches. The convolutional trunk contains three one-dimensional convolutional layers with 300, 200 and 200 filters, with kernel widths of 19, 11 and 7, respectively, and batch normalization in the convolutional network. The shared representation is passed through a fully connected layer with 1,000 hidden units and dropout. The model then branches into separate output heads for K562, HepG2 and SK-N-SH; each branch contains three fully connected layers with 140 hidden units and returns predicted MPRA \(\log_2\) fold-change for one cell type.

For each submitted library, the architecture, optimization procedure and evaluation pipeline were fixed. Training used the Malinois objective inherited from the released implementation, combining an \(\mathrm{L}_1\) activity loss with a \(\mathrm{KL}\)-divergence term. Optimization used Adam with learning rate \(3.27 \times 10^{-3}\), \(\beta = (0.866, 0.879)\), \(\varepsilon = 10^{-8}\), weight decay \(3.44 \times 10^{-4}\) and AMSGrad. The learning-rate schedule was CosineAnnealingWarmRestarts with \(T_0 = 4096\), \(T_{\mathrm{mult}} = 1\) and \(\eta_{\min} = 0\). Models were trained with batch size 512 for up to 200 epochs, with early stopping on a 10\% random validation split.

Forward/reverse-complement averaging was applied only when Malinois generated labels, including labels for submitted libraries and oracle-labeled evaluation sets. Downstream models trained inside MPRAbox were trained on single-strand inputs without reverse-complement augmentation and evaluated using a single forward pass. All oracle inference and downstream model training were run on an NVIDIA DGX Spark.

\subsection{Evaluation suite}\label{evaluation-suite}

Trained models were evaluated on 14 held-out evaluation sets derived from nine source distributions (Table 1). Five sets used empirical MPRA measurements as labels and nine used Malinois-generated labels. The empirical-label sets provide direct evaluation against measured regulatory activity, while the oracle-labeled sets extend the suite to sequence distributions not represented in the experimental data.

The evaluation suite included held-out reporter sequences from the Gosai et al. MPRA experiment,\cite{Gosai2024-lx} UK Biobank and GTEx variant sequences,\cite{Siraj2026-na} DNase I hypersensitive sites from the Meuleman et al. DHS index,\cite{Meuleman2020-fu} regulatory sequence classes assigned by the deep learning model Sei,\cite{Chen2022-np} random genomic windows and fully synthetic random sequences. Chromosome-level holdout was enforced during construction. Evaluation sets drawn from the genome used chromosomes excluded from Malinois training, including chromosomes 7, 13, 19, 21 and X. UK Biobank and GTEx variant sets were constructed as 200 bp windows centered on candidate causal variants. Depending on the evaluation set, either a single allele per locus or both reference and alternate alleles were retained. Evaluation set identities were hidden from the agent. Performance was computed as Pearson correlation between predicted and labeled MPRA \(\log_2\) fold-change across the included sequences.

\subsection{Human-curated baseline strategies}\label{human-curated-baseline-strategies}

We constructed 14 human-curated library-design strategies as a reproducible reference panel. Each strategy defined a sampling rule rather than a fixed library; for each rule, we generated five independent libraries using distinct sampling seeds, so that replicates differed only in the specific sequences drawn under the same design procedure. The strategies sampled from four sequence sources: accessible chromatin regions from the DHS index, regulatory sequence classes from Sei, fully synthetic random DNA, and previously assayed MPRA sequences from the Malinois training set.

DHS-based strategies sampled from the DNase I hypersensitive site index, a genome-wide catalog of accessible chromatin regions.\cite{Meuleman2020-fu,Chen2022-np} The DHS index includes non-negative matrix factorization (NMF) topic annotations that summarize accessibility patterns across biosamples into 16 regulatory programs. We used these annotations to define two structured sampling schemes. In topic-weighted sampling, DHS elements were sampled with probability proportional to their NMF topic loadings. In topic-stratified sampling, each element was assigned to its maximum-loading topic, and equal numbers of sequences were sampled from each of the 16 topics. Uniform DHS sampling treated all DHS elements as equally likely.

Sei-based strategies sampled from genomic regions annotated by Sei regulatory sequence class.\cite{Chen2022-np} Sei regions were sampled either in proportion to class frequency or with equal allocation across regulatory classes. Synthetic random libraries were generated by drawing each nucleotide independently from a uniform distribution over \(\{A,C,G,T\}\). Prior MPRA libraries were sampled uniformly from the Malinois training MPRA sequence set and evaluated either with Malinois-generated labels or with the original empirical measurements.\cite{Gosai2024-lx} The full panel also included two- and three-source mixtures combining DHS, Sei and synthetic sequences. The 14 human-curated design strategies are listed in Table 2.

Strategies were evaluated at library sizes of 10,000, 25,000, 50,000, 100,000, 150,000, 200,000 and 300,000 sequences. The primary comparison point was 50,000 sequences, matching the library size used for agent submissions.

\subsection{Agent configuration and regimes}\label{agent-configuration-and-regimes}

We instantiated three model-harness systems using each provider's native coding-agent environment: Claude Opus 4.7 in Claude Code, GPT-5.5 in Codex CLI, and Gemini 3.5 Flash in Gemini CLI. All three operated in a sandboxed Linux environment with access to a shell, Python, the filesystem, and the internet, with permissions configured to allow autonomous tool use; otherwise, we used the default parameters of the respective provider-native harnesses. We introduced no third-party agent scaffold.

\subsection{Dry oracles with invented rules in MPRAbox}

We constructed 14 invented rules for use as dry oracles in MPRAbox. Each dry oracle replaced Malinois as the function that assigns activity labels to sequences while preserving the same basic library-design task. Given a submitted library of \(N\) 200-bp sequences, the dry oracle assigned three synthetic activity values to each sequence, mirroring the three-cell-type output format of Malinois. These sequence-level values were hidden from the agent. The 14 rules spanned nucleotide composition, short sequence patterns, positional rules, compressibility, number-theoretic relationships, cellular-automaton dynamics, and a hidden cipher that converted nucleotide pairs into English letters (Table 3).

To score a submitted library, we represented each sequence by its normalized 6-mer frequencies and trained a three-output Ridge regression model (\(\alpha = 1.0\)) on the submitted sequences and their oracle-generated labels. We then evaluated this model on the same 14 held-out sequence sets used in MPRAbox, after relabeling those sequences with the same dry oracle. For each evaluation set, we calculated the Pearson correlation between the model predictions and the hidden oracle labels and averaged across the three outputs; we then averaged across the 14 evaluation sets. The agent received only this final aggregate performance score, not the sequence-level oracle labels, the individual output values, or the hidden scoring rule.

For each evaluation set, performance is measured as the Pearson correlation between predicted and oracle-generated labels, averaged across the three output dimensions. For each library, we report the mean across the 14 MPRAbox evaluation sets.

We ran Claude Opus 4.7 on each of the 14 dry oracles under three task framings. In the MPRA-framed condition, the agent was told that it was designing regulatory DNA for an MPRA library. In the unframed condition, the agent still designed sequences over \(\{A,C,G,T\}\), but was told only that it was optimizing against a black-box scoring system, with no reference to MPRA or gene regulation. In the symbolic condition, we replaced the DNA alphabet with \(\{0,1,2,3\}\). The hidden scoring rule was identical across the three framings. We ran each oracle-framing combination independently for 30 rounds.

\subsection{CodonBox implementation}

CodonBox places the agent in invented biological worlds, each governed by a concealed genetic system whose rules are unknown to the agent but yield a measurable fitness readout. On every round the agent (Claude Opus 4.7 in a minimal harness, see \url{https://github.com/asr2210/science-sandbox}) submits one candidate sequence. A sealed oracle translates that sequence under a world-specific hidden rule, folds the resulting product with a fixed physical model, and returns a single fitness number and nothing more. Neither the translation rule nor the folding step is visible to the agent, so it can only probe each world through its input--output behavior and must deduce the mapping from sequence to fitness by experiment alone. Since the true rule of each world is set in advance by us but withheld from the agent, the setup separates agents that actually reconstruct the governing rule from those that only accumulate a list of high-scoring sequences.

The physical substrate is Dill\textquotesingle s \(HP\) model of protein folding,\cite{Dill1985-vb} which reduces a protein to a string of hydrophobic (\(H\)) and polar (\(P\)) residues. Every translated product in CodonBox is a 16-residue \(H/P\) string that folds on a two-dimensional square lattice, and its fitness is the count of favorable non-consecutive \(H\)--\(H\) contacts in the lowest-energy fold. We enumerated this value for all \(2^{16} = 65{,}536\) possible 16-residue \(H/P\) strings ahead of time, which fixes the folding physics while letting the genetic layer differ from world to world. On this lattice no 16-residue chain can exceed 9 such contacts, so the same fitness ceiling of 9 holds in every world.

Each hidden rule defines how an input string is segmented into codons and how each codon becomes an \(H\) or \(P\) residue. Varying these rules gave us eight worlds that differ along four axes: codon length (2, 3, or 4 characters), alphabet size (4, 6, or 8 characters), whether a codon contains silent positions that leave its residue unchanged, and whether the informative positions act additively or interact. A control world mirrors familiar genetics, with a four-character alphabet, three-character codons, and independent contributions from all three positions. Two worlds are built so that changing one character at a time reveals little: in one, only a codon\textquotesingle s central position sets the residue and the two outer positions are silent; in the other, the two outer positions jointly determine the residue through a non-additive interaction while the center is silent. The hardest world layers several of these features together, pairing a six-character alphabet with four-character codons, a silent position, and a three-position joint dependence. Each world\textquotesingle s codon-to-residue table was built deterministically from its parameters under a fixed seed and balanced so that exactly half of the possible codons map to \(H\) and half to \(P\). For each world, we ran Claude Opus 4.7 once, for 500 design rounds.

\clearpage
\backmatter

\bmhead{Data and code availability}\label{data-and-code-availability}

\begingroup
\raggedright
All code used to produce the sandboxes presented here is available at
\mbox{\url{https://github.com/asr2210/science-sandbox}}.\par
\endgroup

\bmhead{Funding}

This work was made possible by funding from the Ladders to Cures Scientific Accelerator of the Broad Institute of MIT and Harvard and the Howard Hughes Medical Institute.

\bmhead{Competing Interests}

P.C.S. holds several patents related to diagnostic technologies and is a cofounder and equity holder in Delve Biosciences and Lyra Labs, a board member and equity holder in Polaris Genomics, and an equity holder of NextGenJane. P.C.S. was formerly a co-founder of Sherlock Biosciences and a board member of Danaher Corporation, until December 2024. S.N.B. reports interests in Amplifyer Bio, Catalio Capital, Danaher, Earli Inc., Impilo Therapeutics, Matrisome Bio, Ochre Bio, Pictet, Port Therapeutics, Ropirio Therapeutics, Satellite Bio, Sunbird Bio, Vertex Pharmaceuticals, and Xilio Therapeutics. S.N.B.'s interests are reviewed and managed under MIT's policies for potential conflicts of interest. S.S. serves on the scientific advisory board of Waypoint Bio, Deepcell, Novo Nordisk, and is an advisor for Dewpoint Therapeutics.

\bmhead{Acknowledgements}

We thank Debora Marks and Robert Langer for insightful discussions that helped shape this study.

\bibliography{references}

\clearpage
\bmhead{Supplementary Figures and Tables}
\setcounter{figure}{0}
\renewcommand{\thefigure}{S\arabic{figure}}
\renewcommand{\theHfigure}{S\arabic{figure}}

\begin{center}
\begin{minipage}{\textwidth}
\centering
\iflatexml
\includegraphics[width=800pt]{figures/html-svg/supplementary_figure1.svg}
\else
\sourcefigure[width=\textwidth,height=0.82\textheight,keepaspectratio]{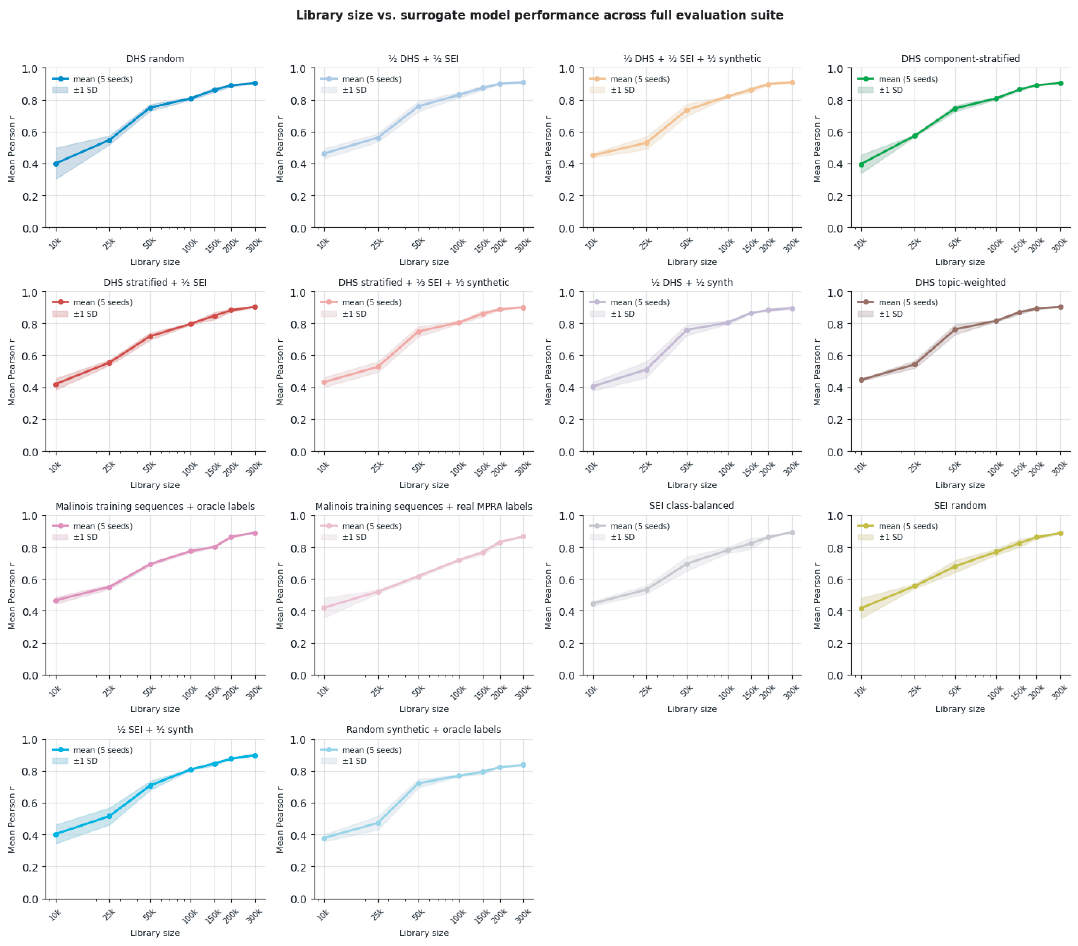}
\fi
\captionof{figure}{\textbf{Performance versus training library size.} Performance against library size (10,000--300,000 sequences) for each of the 14 human-curated strategies (one panel each). Line, mean over five seeds; band, ± 1 SD.}
\label{fig:supp-library-size}
\end{minipage}
\end{center}

\clearpage
\begin{figure}[p]
\centering
\iflatexml
\includegraphics[width=800pt]{figures/html-svg/supplementary_figure2.svg}
\else
\sourcefigure[width=\textwidth]{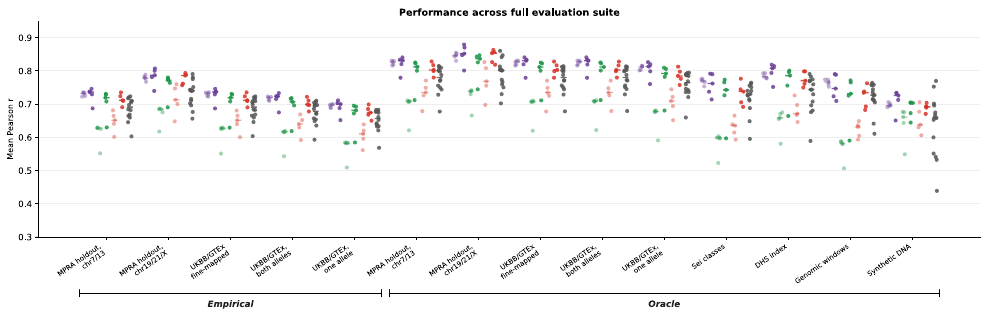}
\fi
\caption{Per-set performance of one-shot libraries by model across each held-out evaluation set, which includes empirical MPRA measurements and oracle-labeled regulatory, variant, genomic, and synthetic sequence sets.}
\label{fig:supp-per-set}
\end{figure}

\clearpage
\begin{figure}[p]
\centering
\iflatexml
\includegraphics[width=800pt]{figures/html-svg/supplementary_figure3.svg}
\else
\sourcefigure[width=\textwidth]{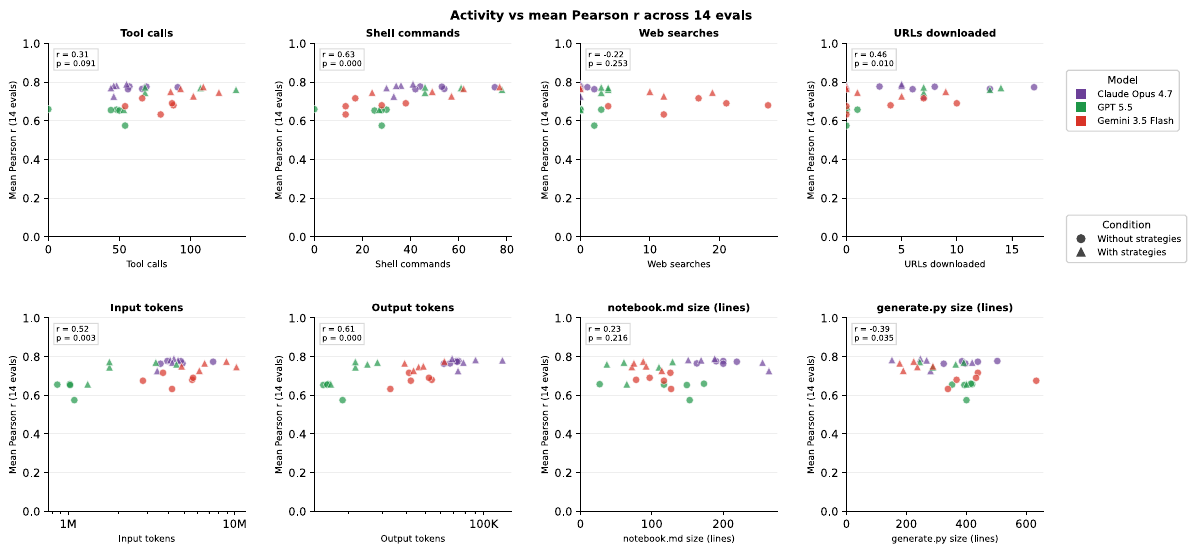}
\fi
\caption{\textbf{Agent activity versus one-shot performance.} One-shot performance against eight activity metrics (points colored by model). Each panel shows the Pearson \(r\) and two-sided \(p\) across the 30 runs.}
\label{fig:supp-activity}
\end{figure}

\clearpage
\begin{figure}[p]
\centering
\iflatexml
\includegraphics[width=800pt]{figures/html-svg/supplementary_figure4.svg}
\else
\sourcefigure[width=\textwidth]{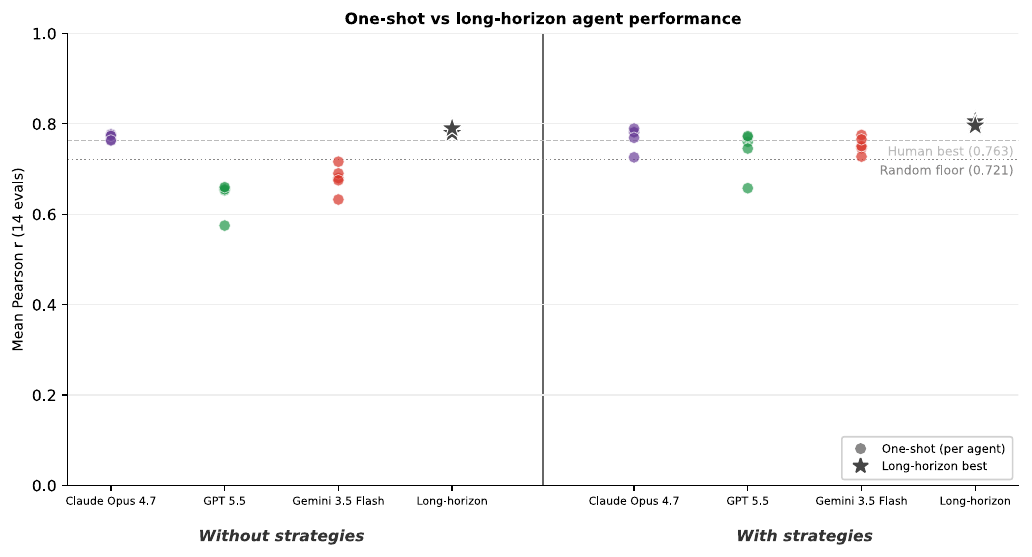}
\fi
\caption{\textbf{One-shot versus long-horizon performance.} One-shot performance (points, by model) and the best long-horizon design (star) in the blind and informed conditions. Dashed line, human best; dotted line, random floor.}
\label{fig:supp-one-shot-long-horizon}
\end{figure}

\clearpage
\begin{figure}[p]
\centering
\iflatexml
\includegraphics[width=800pt]{figures/html-svg/supplementary_figure5.svg}
\else
\sourcefigure[width=\textwidth,height=0.80\textheight,keepaspectratio]{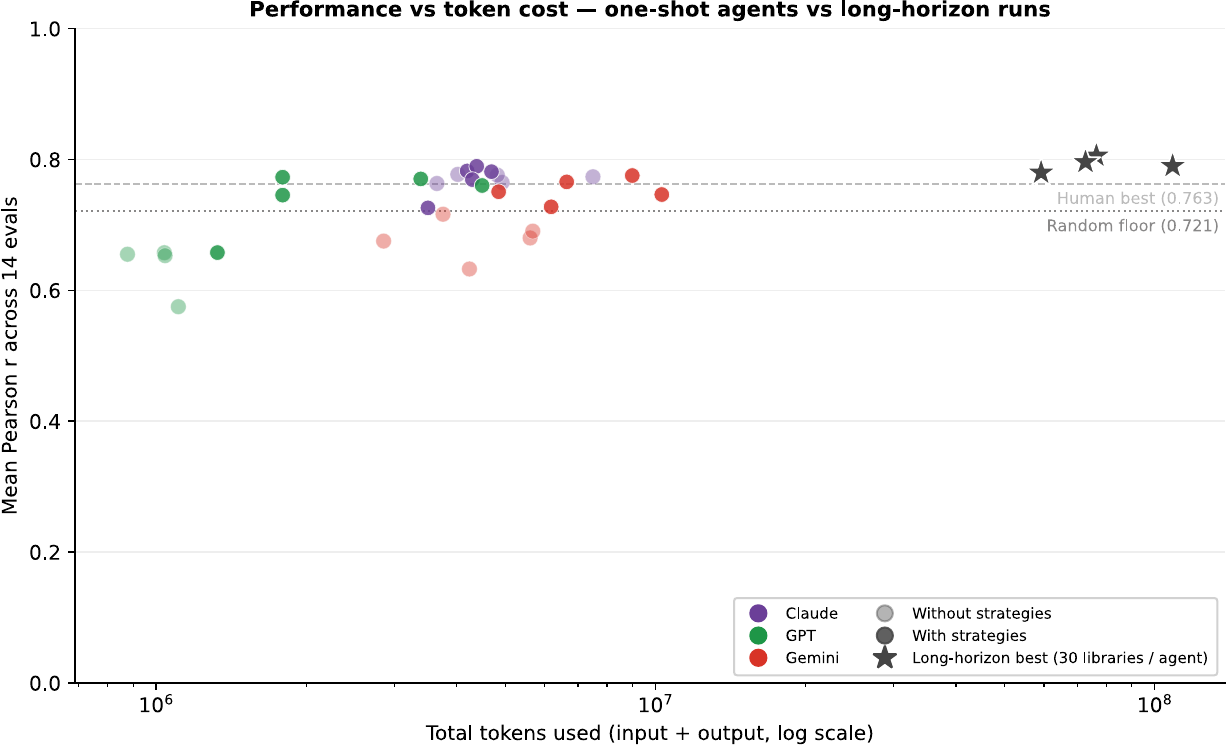}
\fi
\caption{\textbf{Performance versus token cost.} Performance against total tokens used (input + output, log scale) for one-shot agents (points, by model) and long-horizon runs (stars, best of 30 libraries per agent).}
\label{fig:supp-token-cost}
\end{figure}

\clearpage
\bmhead{Supplementary Agent Instructions and Harness}

All agent-facing files in both sandboxes are provided directly below.

\supplementarysource[]{MPRAbox one-shot instructions: without prior knowledge}{mprabox/instructions/oneshot_without_prior_knowledge.md}{science-sandbox-current/mprabox/instructions/oneshot_without_prior_knowledge.md}

\supplementarysource{MPRAbox one-shot instructions: with prior knowledge}{mprabox/instructions/oneshot_with_prior_knowledge.md}{science-sandbox-current/mprabox/instructions/oneshot_with_prior_knowledge.md}

\supplementarysource{MPRAbox informed-condition strategy reference}{mprabox/instructions/strategies.md}{science-sandbox-current/mprabox/instructions/strategies.md}

\supplementarysource{MPRAbox long-horizon instructions: without prior knowledge}{mprabox/instructions/long_horizon_without_prior_knowledge.md}{science-sandbox-current/mprabox/instructions/long_horizon_without_prior_knowledge.md}

\supplementarysource{MPRAbox long-horizon instructions: with prior knowledge}{mprabox/instructions/long_horizon_with_prior_knowledge.md}{science-sandbox-current/mprabox/instructions/long_horizon_with_prior_knowledge.md}

\supplementarysource{MPRAbox dry-oracle instructions: MPRA-framed}{mprabox/dry_oracles/instructions/mpra_framed.md}{science-sandbox-current/mprabox/dry_oracles/instructions/mpra_framed.md}

\supplementarysource{MPRAbox dry-oracle instructions: unframed}{mprabox/dry_oracles/instructions/unframed.md}{science-sandbox-current/mprabox/dry_oracles/instructions/unframed.md}

\supplementarysource{MPRAbox dry-oracle instructions: symbolic}{mprabox/dry_oracles/instructions/symbolic.md}{science-sandbox-current/mprabox/dry_oracles/instructions/symbolic.md}

\supplementarysource{CodonBox instructions}{codonbox/instructions.md}{science-sandbox-current/codonbox/instructions.md}

\begingroup
\renewcommand{\sourcebaselinestretch}{0.93}
\supplementarysource{CodonBox harness}{codonbox/harness.py}{science-sandbox-current/codonbox/harness.py}
\endgroup

\end{document}